\RequirePackage{ifpdf}
\documentclass[letterpaper]{JHEP3}
\usepackage{epsfig}
\usepackage{comment}
\usepackage{amsmath}

\def\ba{\begin{eqnarray}}
\def\ea{\end{eqnarray}}
\def\be{\begin{equation}}
\def\ee{\end{equation}}
\def\nn{\nonumber}
\def\exd{{\rm d}}

\makeatletter
\def\x@arrow{\DOTSB\Relbar}
\def\xlongequalsignfill@{\arrowfill@\x@arrow\Relbar\x@arrow}
\newcommand{\xlongequal}[2]{%
    \ext@arrow 0099\xlongequalsignfill@{#1}{#2}}
\makeatother

\newcommand{\roughly}[1]{\mathrel{\raise.3ex\hbox{$#1$\kern-0.85em
\lower1ex\hbox{$\sim$}}}}

\def\nott#1{\setbox0=\hbox{$#1$}                
   \dimen0=\wd0                                 
   \setbox1=\hbox{/} \dimen1=\wd1               
   \ifdim\dimen0>\dimen1                        
      \rlap{\hbox to \dimen0{\hfil/\hfil}}      
      #1                                        
   \else                                        
      \rlap{\hbox to \dimen1{\hfil$#1$\hfil}}   
      /                                         
   \fi}                                         %

\def\endignore{}
\def\ignore #1\endignore{} 

\def\be{\begin{equation}}
\def\beq\begin{equation}
\def\ee{\end{equation}}
\def\bea{\begin{eqnarray}}
\def\eea{\end{eqnarray}}

\def\ket{\rangle}

\def\eqa{\begin{eqnarray}}
\def\eeqa{\end{eqnarray}}
\def\eq{\begin{equation}}
\def\eeq{\end{equation}}

\def\nn{\nonumber}

\def\pref#1{(\ref{#1})}

\def\exd{{\rm d}}
\def\nn{\nonumber}
\def\pref#1{(\ref{#1})}
\def\be{\begin{equation}}
\def\ee{\end{equation}}

\def\beq{\begin{equation}}
\def\eeq{\end{equation}}
\def\beqa{\begin{eqnarray}}
\def\eeqa{\end{eqnarray}}

\def\cA{{\cal A}}

\def\cC{{\cal C}}
\def\cD{{\cal D}}

\def\cG{{\cal G}}
\def\cH{{\cal H}}

\def\cJ{{\cal J}}

\def\cL{{\cal L}}

\def\cO{{\cal O}}

\def\cR{{\cal R}}
\def\cS{{\cal S}}
\def\cT{{\cal T}}

\def\cV{{\cal V}}

\def\ssA{{\scriptscriptstyle A}}
\def\ssB{{\scriptscriptstyle B}}

\def\ssD{{\scriptscriptstyle D}}

\def\ssF{{\scriptscriptstyle F}}

\def\ssM{{\scriptscriptstyle M}}
\def\ssN{{\scriptscriptstyle N}}
\def\ssP{{\scriptscriptstyle P}}
\def\ssQ{{\scriptscriptstyle Q}}
\def\ssR{{\scriptscriptstyle R}}

\def\ssW{{\scriptscriptstyle W}}

\def\ssZ{{\scriptscriptstyle Z}}

\def\KK{{\scriptscriptstyle KK}}

\def\SN{{\scriptscriptstyle SN}}
\def\QCD{{\scriptscriptstyle QCD}}

\def\EW{{\scriptscriptstyle EW}}

\def\bfx{{\bf x}}

\newcommand{\bmat}{\left(\begin{array}}
\newcommand{\emat}{\end{array}\right)}

\def\-{\hphantom{-}}

\def\s2{\frac{1}{2}}

\def\IF{\relax{\rm I\kern-.18em F}}
\def\II{\relax{\rm I\kern-.18em I}}
\def\IP{\relax{\rm I\kern-.18em P}}
\def\IC{\relax{\rm I\kern-.48em C}}
\def\IR{\relax{\rm I\kern-.18em R}}
\def\IK{\relax{\rm I\kern-.20em K}}
\def\IM{\relax{\rm I\kern-.25em M}}

\def\y2{Y_{\ssM\ssN} Y^{\ssM\ssN}}
\def\Riem2{R_{\ssA\ssB\ssM\ssN} R^{\ssA\ssB\ssM\ssN}}
\def\Ricci2{R_{\ssM\ssN} R^{\ssM\ssN}}

\def\f2{F^{a}_{\ssM\ssN} F^{\ssM\ssN}_a}

\def\Asl{\hbox{/\kern-.7500em\it A}} 
\def\dsl{\hbox{/\kern-.5500em$\partial$}}
\def\pxpsl{\hbox{/\kern-.5600em$p$}}
\def\Dsl{\,\raise.15ex\hbox{/}\mkern-13.5mu D}
\def \one{\relax{\rm 1\kern-.26em I}}

\def\exd{{\rm d}}

\def\nn{\nonumber}

\def\({\left(}
\def\){\right)}

\preprint{PRELIMINARY DRAFT}

\title{The Cosmological Constant Problem:\\
Why it's hard to get Dark Energy from Micro-physics\footnote{Lectures given to the Les Houches Summer School ``Post-Planck Cosmology,'' 8 July - 2 August 2013}}

\author{C.P. Burgess\\

Department of Physics \& Astronomy, McMaster University,\\
\qquad 1280 Main Street West, Hamilton, Ontario, Canada, L8S 4M1.\\
Perimeter Institute for Theoretical Physics,\\ \qquad 31 Caroline
Street North, Waterloo, Ontario, Canada, N2L 2Y5. }

\date{}

\abstract{These notes present a brief introduction to `naturalness' problems in cosmology, and to the Cosmological Constant Problem in particular. The main focus is the `old' cosmological constant problem, though the more recent variants are also briefly discussed. Several notions of naturalness are defined, including the closely related ideas of technical naturalness and `t Hooft naturalness, and it is shown why these naturally arise when cosmology is embedded within a framework --- effective field theories --- that efficiently captures what is consistent with what is known about the physics of smaller distances. Some care is taken to clarify conceptual issues, such as the relevance or not of quadratic divergences, about which some confusion has arisen over the years. A set of minimal criteria are formulated against which proposed solutions to the problem can be judged, and a brief overview made of the general limitations of most of the approaches. A somewhat more in-depth discussion is provided of what I view as the most promising approach. These notes are aimed at graduate students with a basic working knowledge of quantum field theory and cosmology, but with no detailed knowledge of particle physics.}

\begin{document}


\section{The Problem}
\label{sec:problem}

What is the vacuum's energy, and how does it gravitate? This deceptively simple question is relevant to our present understanding of cosmology \cite{PlanckParams} but also cuts to the core of one of the deepest flaws to our otherwise sophisticated understanding of how Nature works. It is relevant to cosmology because a vacuum energy would affect the way the universe expands in an observable way. Yet it is a problem because observations indicate a vacuum energy that is more than 50 orders of magnitude smaller than we think it should be, given the particles and interactions that we find in experiments.

Most disturbingly, as these notes argue at some length, this mismatch between prediction and observations is not so easily solved. Normally such a large disagreement with data immediately spawns many new candidate theories, designed to remove the discrepancy without ruining the success of other experimental tests. More experiments are then required to decide which of the new proposals is the one chosen by Nature. The same is {\em not} true for the vacuum energy, for which there is not even one candidate modification of theory which all agree can provide a potentially viable way to reconcile observations with predictions \cite{WbgNoGo, CCrevs}.

As we shall see, the problem arises because of the robustness of the prediction of a large vacuum energy, together with the apparent simplicity of the argument of what its gravitational consequences must be. These two seem to follow almost directly as a consequence of one of the great successes of 20th century physics: the emergence of quantum field theory as the language in which the laws of Nature appear to be written. Quantum field theory provides the class of theories that reconcile the somewhat contradictory requirements of special relativity and quantum mechanics. Evidence for its success in describing Nature first emerged convincingly with the exquisitely precise tests of Quantum Electrodynamics, followed later by the extension to the other non-gravitational interactions in the form of the Standard Model of particle physics. Quantum field theory also appears to be the right framework for understanding gravitational physics, including the quantum corrections to the classical predictions of General Relativity, at least when restricted to the long-distance observables to which we have direct experimental access (including in particular in cosmology) \cite{GREFT}.

The great generality of the problem has led some to argue that something very fundamental must be breaking down, possibly the applicability of quantum field theory itself. These notes close with an argument that things may not be as bad as this, and that there may be a less exotic way out. And if there is, it is a very important clue. The very absence of many competing models means that if a successful one is found, this is likely the way Nature works.

And this ultimately is also why pragmatic people like astronomers and observational cosmologists should care about the cosmological constant problem. After all, a large class of theories has been devised to describe dark energy in cosmology, and early-universe observations are unlikely in themselves to be sufficient to distinguish which is the right one. But we need not only use just these observations, since doing so would ignore another important class of clues about the nature of dark energy. These other clues are the constraints that arise from requiring the theories that describe cosmology be obtainable as the long-distance limit of a sensible theory of all of the rest of what we know about more microscopic physics.

Indeed cosmology appears to provide one of the few loopholes to the general observation that Nature decouples; that is, that the details of short-distance physics do not matter much for describing physics on much larger distance scales. (For instance one does not have to be an expert in nuclear physics to understand the properties of atomic energy levels, and this property of Nature --- that each scale can be understood largely on its own terms --- ultimately underlies why progress in science is possible at all.) Cosmology is special because the things that seem needed to describe observations --- things like a vacuum energy or a very light scalar field --- are among the few kinds of things that {\em are} sensitive to the details of small-distance physics. So it behooves us to take advantage of this sensitivity to constrain the class of cosmological models we should entertain, perhaps thereby arriving at a sufficiently small class of theories that cosmological observations might suffice to distinguish which one Nature has chosen.

The remainder of these notes are organized as follows. The section you are now reading, \S\ref{sec:problem}, reviews precisely what the cosmological constant problem is. This starts in \S\ref{ssec:vacgrav} with a discussion of the constraints on how the vacuum energy would gravitate that follow from Lorentz invariance. This is followed in \S\ref{ssec:cosmoexpl} by a brief summary of the evidence for dark energy and the little that is known about it. \S\ref{ssec:naivecc} then sets up the naive, traditional formulation of the `old' cosmological constant problem in terms of ultraviolet divergences, and then criticizes this formulation as being needlessly misleading. A better formulation of the problem is then given in \S\ref{ssec:MoreEffectiveOld}, in terms of effective field theories. Given this formulation \S\ref{ssec:Criteria} proposes three minimal criteria that any given proposal must pass to succeed in solving the problem. Finally, \S\ref{ssec:othercc} formulates the various `new' cosmological constant problems that arose because of the discovery of a nonzero dark-energy density.

A few approaches to solving the `old' problem are given in \S\ref{section:approaches}, starting with a telegraphic summary of several of the most commonly found ones, in \S\ref{ssec:roadstrav}, and ending in \S\ref{ssec:forward} with a more detailed description of the approach I believe to provide the most hopeful tack at present. Some conclusions are summarized in the final section, \S\ref{sec:summary}.

\subsection{How would the vacuum gravitate?}
\label{ssec:vacgrav}

If the vacuum did have an energy, how would it gravitate? Although there is much we don't know about the vacuum, one thing we {\em do} know is that it appears to be Lorentz invariant. That is, it appears to be invariant under arbitrary rotations of the three axes of space and it seems to be identical as seen by arbitrary observers moving relative to one another at constant speed.

Evidence for these properties comes from a variety of experimental tests. These range from the classic Michelson-Morley experiments of how light rays propagate; the independence of atomic properties measured in the lab from the Earth's motion in its orbit around the Sun; and so on \cite{LIvactests}. Some of these --- the stability of atomic properties, for instance --- are very accurate; in some cases constraining deviations from Lorentz invariance to one part in $10^{20}$ or more.

Lorentz invariance tells us much about how the vacuum would gravitate if it were to have a nonzero energy density, $\rho_{\rm vac}$. The only stress energy that is consistent with Lorentz invariance is\footnote{These notes use metric signature $(-+++)$ --- as should you --- and Weinberg's curvature conventions \cite{Wbg} (which differ from MTW's \cite{MTW} only in the overall sign of the Riemann tensor).}
\be \label{vacTmn}
 T^{\mu\nu}_{\rm vac} = -\rho_{\rm vac} \, g^{\mu\nu} \,,
\ee
where $g^{\mu\nu}$ is the inverse spacetime metric. Furthermore, given this form stress energy conservation, $\nabla_\mu T^{\mu\nu} = 0$, requires $\rho_{\rm vac}$ must be a constant. Eq.~\pref{vacTmn} looks like a special case of the stress-energy of a perfect fluid, which would be
\be \label{fluidT}
 T^{\mu\nu} = p \, g^{\mu\nu} + (p + \rho) \, u^\mu u^\nu \,,
\ee
where $p$ is the pressure, $\rho$ the energy density and $u^\mu$ is the 4-velocity of the fluid. The vacuum is the special case where the pressure satisfies\footnote{Fundamental units with $\hbar = c = k_\ssB = 1$ are used everywhere, unless explicitly stated otherwise.} $p_{\rm vac} = - \rho_{\rm vac}$ and so its equation of state parameter is
\be
 w := \frac{ p_{\rm vac}}{\rho_{\rm vac}} = -1 \,.
\ee
Within the context of cosmology this is the pressure that is required to do the work necessary to ensure that a constant energy density can be consistent with the expansion of space.

A stress energy of the form of eq.~\pref{vacTmn} is very familiar when inserted into Einstein's equations,
\be \label{Einstein}
 R^{\mu\nu} - \frac12 \, R \, g^{\mu\nu} + \kappa^2 \, T^{\mu\nu} = 0 \,,
\ee
where $\kappa^2 = 8 \pi G_\ssN$ with $G_\ssN$ being Newton's constant. It plays the role of Einstein's cosmological constant,
\be \label{EinsteinCC}
 R^{\mu\nu} - \frac12 \, R \, g^{\mu\nu} - \lambda \, g^{\mu\nu} = 0 \,,
\ee
with $\lambda = \kappa^2 \rho_{\rm vac}$.

The presence of such a term is an obstruction to having small curvatures, inasmuch as it implies
\be \label{RicciCC}
 R^{\mu\nu} + \kappa^2 \left( T^{\mu\nu} - \frac12 \, T \, g^{\mu\nu} \right) = R^{\mu\nu} + \kappa^2 \rho_{\rm vac} g^{\mu\nu} = 0 \,,
\ee
where $T := g_{\lambda\rho} T^{\lambda\rho}$. As applied to a spatially flat expanding Friedmann-Lemaitre-Robertson-Walker cosmology, with metric
\be
 \exd s^2 = - \exd t^2 + a^2(t) \, \exd \bfx \cdot \exd \bfx \,,
\ee
the case $\rho_{\rm vac} > 0$ implies an accelerating expansion,
\be
 a(t) = a_0 \, e^{H (t-t_0)} \quad \hbox{with} \quad H^2 = \frac13 \, \kappa^2 \rho_{\rm vac} \,.
\ee
What is remarkable about modern cosmology is that there is now evidence for this kind of acceleration in the universe's recent history.

\subsection{Cosmology: What must be explained?}
\label{ssec:cosmoexpl}

Detailed measurements of the properties of the Cosmic Microwave Background (CMB) have allowed the Hot Big Bang model of cosmology to be tested with unprecedented redundancy and precision, and it has emerged all the stronger for having done so. The redundancy of these tests gives confidence that the basic picture --- the expansion of an initial hot primordial soup --- is basically right. Their precision allows a detailed inference of the model's parameters, including the first-ever survey of the energy content of the Universe viewed as a whole.

\FIGURE[tbh]{
\begin{tabular}{ll}
\hspace{-1.9cm}\epsfig{file=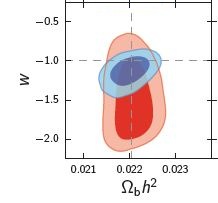,height=0.345\hsize} &
\epsfig{file=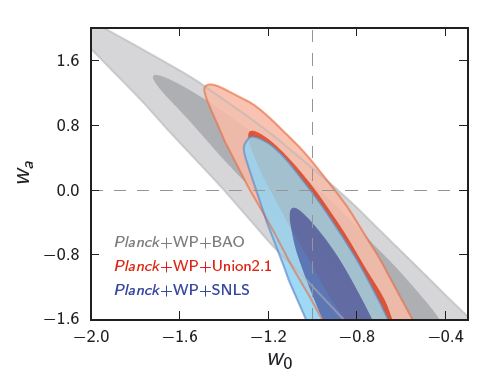,height=0.345\hsize}
\end{tabular}
\caption{Left panel: Planck constraints on the vacuum equation of state parameter $w$. Right panel: Planck constraints on the cosmological rate of change of $w$. Both figures taken from \cite{PlanckParams}\label{Planckwfig}
}}

The biggest surprise to emerge from these tests is the requirement for the existence of {\em two} --- not just one --- types of unknown constituents, whose presence is inferred from the way they gravitate.\footnote{Since gravity is used to infer their existence, many have suggested modifying gravity instead of proposing two new kinds of matter. Unfortunately, no convincing modification has yet been found although the search is actively being pursued. But even if found, any such a modification must avoid the cosmological constant problem as discussed below.} One of these --- dark matter --- is required by a variety of astrophysical and cosmological measurements, and appears to make up around a quarter of the observed total energy density. Whatever it is, given how its presence is inferred it must cluster under gravitational attraction and so cannot cause an acceleration of the universal expansion. Because of this it is the other unknown component --- dark energy --- that is of more interest for the present purposes.

The evidence for dark energy, which standard cosmology says makes up some 70\% of the total energy density, comes from two main sources. First, the CMB provides evidence that the universe is spatially flat \cite{PlanckParams}, in which case Einstein's equations imply the total energy density must be critical, and so in particular contain more than just ordinary matter and dark matter. Second, tests of the distance-redshift relation using distant supernovae indicate that the universal expansion is accelerating \cite{Accelerating}. Furthermore, the amount of acceleration is consistent with having a spatially flat universe if the dark energy fluid has an equation of state parameter consistent with $w = -1$ (see, for instance, Fig. \ref{Planckwfig}). Translated into an energy density, the observation that dark energy makes up 70\% of the critical density implies that the vacuum energy density would be
\be \label{rhovobs}
 \rho_{\rm vac}(\hbox{observed}) \simeq (3 \times 10^{-3} \; \hbox{eV})^4 \,.
\ee

Of course, observations in themselves do not directly tell us that the dark energy must be a vacuum energy (or cosmological constant). Indeed many models have been built that relax the lorentz invariance assumption, such as by involving slowly varying scalar fields, $\phi^i$. If the kinetic energy density for these fields is $K = \frac12 \, \cG_{ij}(\phi) \partial_t \phi^i \partial_t \phi^j$ and their scalar potential is $V(\phi)$, then their total energy density and pressure work out to be $\rho = K + V$ and $p = K - V$, and so $w = p/\rho$ is a monotonically increasing function of $K/V$ that satisfies $-1 \le w \le 1$, and is close to $-1$ whenever $V$ is much bigger than $K$. Since this requires $\phi^i$ to vary slowly, it requires the potential to have a very slowly varying region along which it is close to constant. Such a region becomes observationally indistinguishable from a cosmological constant in the limit that the potential is perfectly flat, and the fields sit motionless in field space with $K = 0$. This generalization appears to open up a large class of theories (and even these do not by any means exhaust the possibilities) from which a description of dark energy might be found.

But from the point of view of observations, the more complicated scalar models are not required, and the dark energy looks exactly like what would be predicted by a cosmological constant. Either way, there seems to be no shortage of models with which to compare observations, of which the simplest requires us to believe nothing more than that the vacuum has a nonzero energy density. So what is the problem?

\subsection{The Naive `Old' CC problem}
\label{ssec:naivecc}

The problem is not that the vacuum energy has been found; there was already a problem when it was thought to be unobservably small. The problem is that we believe the vacuum energy can be computed, and the result should be enormous compared with what has been measured. Because this was already a problem before the dark energy was discovered, this is sometimes called the `old' cosmological constant problem. This distinguishes it from several `new' problems, briefly summarized in the next section, that are more to do with understanding the precise value that has been observed.

There is a naive version of this problem, that is easy to state but is really something of a straw man. This section summarizes this naive formulation, both to highlight its limitations and to motivate the more precise formulation given in the remainder of these notes.

Both the naive and more precise versions of the `old' problem start with a calculation of the energy of the ground state for a quantum field. So consider for these purposes a massive scalar field, $h$. This could be regarded as a physical scalar, such as the Higgs boson, but more generally it can be regarded as a proxy for any of the fields describing one of the elementary particles in your favourite theory of Nature (such as the Standard Model).

To be concrete take the action for the scalar to be
\be \label{ScalarS}
 S = - \int \exd^4 x \sqrt{-g} \; \left[ V(h) + \frac12 \, g^{\mu\nu} \partial_\mu h \, \partial_\nu h \right] \,,
\ee
where the scalar potential could be
\be \label{ScalarV}
 V(h) = V_0 + \frac12 \, V_2 \, h^2 + \frac14 \, V_4 \, h^4 \,,
\ee
with real constants $V_0$, $V_2$, $V_4$ and so on.

If we specialize to a static metric, $\exd s^2 = - \exd t^2 + g_{ij} \, \exd x^i \, \exd x^j$, then time-translation invariance implies there will be a conserved energy.\footnote{The discussion to follow also applies to cosmological situations, where $g_{ij}$ is not time-independent, provided we remain in the adiabatic regime for which the Hubble scale $H$ is much smaller than the mass of the particle of interest.} The energy density (or Hamiltonian density) for the action \pref{ScalarS} is
\be \label{ScalarE}
 \cH = \sqrt{-g} \; \left[ V(h) + \frac12 \, (\partial_t h)^2 + \frac12 \, g^{ij} \partial_i h \, \partial_j h \right] \,,
\ee
and so is bounded below provided $V$ is. Here $i$ and $j$ run over only spatial indices.

In practice, it is often the case that the particle of interest is weakly interacting which allows the vacuum energy to be computed semi-classically.\footnote{For instance if $V_4 \ll 1$ then re-scaling $h \to h/\sqrt{V_4}$ shows that all of the $h$-dependent terms in the action are proportional to $1/V_4$, which means that $V_4$ appears in precisely the same way as does $\hbar$ in the combination $S/\hbar$. This is why small $V_4$ is precisely the same as small $\hbar/S$, and so corresponds to a semi-classical limit.} In the present example this corresponds to expanding the action about its lowest-energy classical solution. Inspection of eq.~\pref{ScalarE} shows that $\cH$ is minimized for constant fields, $\partial_t \bar h = \partial_i \bar h = 0$, with $\bar h$ chosen to minimize $V$: %
\be
 V'(\bar h) = \left( V_2 + V_4 \bar h^2 \right) \bar h = 0 \,.
\ee
For $V_2 \ge 0$ this implies the energy is minimized at $\cH(\bar h) = V_0$ when $\bar h = 0$ while the minimum is instead at $\bar h^2 = - V_2/V_4$ if $V_2 < 0$, at which point the energy density is
\be
 \cH (\bar h) = V_0 - \frac{(V_2)^2}{4 V_4} \,.
\ee
This is the classical vacuum for which we next compute quantum corrections.

The leading quantum correction corresponds to expanding the action to quadratic order in $\hat h$ where (in the interaction picture)
\be
 \hat h(x) := h(x) - \bar h = \sum_k \Bigl[ u_k(x) \, a_k + u_k^*(x) \, a^*_k \Bigr] \,,
\ee
where
\be \label{creannalg}
 [a_k, a_l^*] = \delta_{kl} \,,
\ee
are creation and annihilation operators and $u_k(x)$ are the normalized basis of energy eigenfunctions satisfying the linearized field equation $\Bigl[ -\Box + V''(\bar h) \Bigr] u_k = 0$ and\footnote{It is energy conservation that allows these two operators to be simultaneously diagonalized, and so relies on the background metric being static (or varying with time only adiabatically).} $\partial_t u_k = -i \omega_k \, u_k$ with $\omega_k > 0$.\footnote{If $\omega_k < 0$ for some modes, the Hamiltonian is not bounded from below and cosmology is the least of your problems.} Subleading corrections are then obtained by perturbing this free field with the non-quadratic terms in the action.

In terms of these modes the total energy takes the familiar harmonic-oscillator form
\bea
 \hat H &=& \int \exd^3x \; \cH = \int \exd^3x \sqrt{-g} \; V(\bar h) + \frac12 \sum_k \omega_k \, \left( a_k a^*_k + a^*_k a_k \right) \nn\\
 &=& E_0 + \sum_k \omega_k \, a_k^* a_k \,.
\eea
The last line uses eq.~\pref{creannalg} in the form $\frac12 ( a_k a_k^* + a_k^* a_k ) = a_k^* a_k + \frac12$ and defines
\be
 E_0 = \frac12 \sum_k \omega_k + V(\bar h) \cV \,,
\ee
where $\cV := \int \exd^3 x \sqrt{-g}$ is the total volume of space.\footnote{We can imagine using periodic boundary conditions here and taking $\cV \to \infty$ at the end of the day, though if we do so we must also convert the sum over discrete labels $k$ into an integral over a continuous variable.}

Since this is a set of harmonic oscillators, $\hat H$ can be explicitly diagonalized. The energy eigenstates are obtained by acting repeatedly with $a^*_k$ (corresponding to the creation of particles) on the ground state, $|\Omega\ket$, that is defined by the condition $a_k \, |\Omega \ket = 0$. Applying $\hat H$ to the generic state, $|\Psi\ket = \prod_k \left( a_k^* \right)^{n_k} | \Omega \ket$ with $n_k = 0,1,2,\dots$, we find $\hat H |\Psi \ket = E_\Psi |\Psi \ket$, with eigenvalue
\be
 E_\Psi = E_0 + \sum_k n_k \, \omega_k \,.
\ee

The vacuum is the eigenstate with the lowest energy, corresponding to $|\Omega\ket$, for which $n_k = 0$ $\forall k$. Its energy therefore is $E_0$, and so the prediction for $\hat h$'s contribution to the vacuum energy density is
\be \label{rhopredict}
 \rho_{\rm vac} = \frac{E_0}{\cV} = V(\bar h) + \frac{1}{2\cV} \sum_k \omega_k \,.
\ee
The first term on the right-hand side is the classical energy density, and the second represents the first quantum correction to it.

So far so good. Now comes the controversial part.

The controversial part starts with the observation that the sum over $k$ in eq.~\pref{rhopredict} diverges due to the states with large $\omega_k$ ({\em i.e.} the ultraviolet).\footnote{For infinite volume systems with $k$ a continuous variable, the integration over $k$ actually diverges in two ways: both in the infrared and in the ultraviolet. As we shall see, the infrared divergence is a reflection of the sum being proportional to $\cV$, as is required to have a nonzero contribution to $\rho_{\rm vac}$ in the $\cV \to\infty$ limit.} For instance, in flat space $k$ is simply 3-momentum, and $\omega_k = \sqrt{k^2 + m^2}$ where $m^2 = V''(\bar h)$ is the particle mass, and $(2\pi)^3 \sum_k \omega_k = \cV \int \exd^3 k \, \omega_k$, which diverges quartically as $k \to \infty$. If the integration is cut off so $k \le \Lambda \gg m$, then
\be \label{rhovactheory0}
 \rho_{\rm vac} = V(\bar h) + c_0 \Lambda^4 + c_2 \Lambda^2 m^2 + c_4 m^4 + \cdots \,,
\ee
with calculable dimensionless coefficients $c_i$.\footnote{In order of magnitude typically $c_i$ is an order unity number divided by $(4\pi)^2$.} To this we must add a similar contribution from all other particles in Nature, leading to
\be \label{rhovactheory}
 \rho_{\rm vac} = V(\bar h) + \sum_a \left( c_{0a} \Lambda^4 + c_{2a} \Lambda^2 m_a^2 + c_{4a} m_a^4 + \cdots \right) \,.
\ee
Although the sum over particle types changes the precise coefficients of the different powers of $\Lambda$, it in general does not change the powers of $\Lambda$ that appear.\footnote{Unless of course, the sum over particle types should vanish, such as occurs in supersymmetric theories where the contributions of the bosons can cancel the contributions of the fermions. More about this later.}

At this point what is often said is that we should choose for $\Lambda$ the largest scale about which we think this kind of calculation would be reasonable. This might be taken to be the Planck mass, $M_p := G_\ssN^{-1/2} \simeq 10^{19}$ GeV, since above this scale a full theory of quantum gravity is required, and for which simple quantum field theory could be inadequate in a way that may make the result less sensitive to still-smaller distances. If so, then the mismatch between predictions and observations would be stunningly large:
\be
 \frac{\rho_{\rm vac}(\hbox{theory})}{\rho_{\rm vac}(\hbox{observed})} \simeq \frac{1}{(4\pi)^2} \left( \frac{10^{19} \; \hbox{GeV}}{ 10^{-3} \; \hbox{eV}} \right)^4 \simeq 10^{122} \,.
\ee

But there are several fishy things about this statement. Most importantly, as eq.~\pref{rhovactheory} explicitly shows, the quantum contribution to the vacuum energy is only part of the whole result. In particular, there is also a classical contribution involving unknown constants like $V_0$. How do we know that this doesn't cancel the large quantum contributions to give the small observed value?

How likely is it that parameters like $V_0$ would depend on $\Lambda$ in precisely the right way as to cancel the quantum contributions? Pretty likely, as it turns out. After all, the dependence on $\Lambda$ is a divergent dependence on an artificial scale that is introduced just to regularize some integrals, and in quantum field theory similar divergent dependence also appears in other things, including in the predictions of Quantum Electrodynamics which so famously agree with measurements with exquisite precision. In these other, better understood, cases the divergent dependence on the cutoff is cancelled by renormalization; that is, by a precisely opposite $\Lambda$-dependence of the parameters of the lagrangian, like the electron mass and charge.

So why should we not expect the large quantum correction to the vacuum energy to be similarly absorbed into a renormalization of $V_0$? The short answer is that we should. But, as subsequent sections argue in more detail, this doesn't yet remove the problem. Instead, eq.~\pref{rhovactheory} gets changed to an expression like
\be \label{rhovactheoryR}
 \rho_{\rm vac} = V_\ssR(\bar h) + \sum_a \left( C_{a} m_a^4  + \cdots \right) \,.
\ee
where $V_\ssR$ denotes the {\em renormalized} classical potential, $m_a$ are the renormalized (physical) masses of the particles of interest, and the $C_a$ are calculable, $\Lambda$-independent, coefficients (that are again of order $1/(4\pi)^2$. Although this is numerically smaller than before, each of the terms is still much larger than the observed dark energy density. For instance for the electron alone we have $m_e \simeq 5 \times 10^5$ eV, while for the $Z$ boson we have $m_\ssZ \simeq 90$ GeV, and so these contribute
\bea
 \frac{\rho_{\rm vac}(\hbox{electron})}{\rho_{\rm vac}(\hbox{observed})} &\simeq& \frac{1}{(4\pi)^2} \left( \frac{10^{5} \; \hbox{eV}}{ 10^{-3} \; \hbox{eV}} \right)^4 \simeq 10^{30} \nn\\
\quad \hbox{and} \quad
 \frac{\rho_{\rm vac}(\hbox{$Z$ boson})}{\rho_{\rm vac}(\hbox{observed})} &\simeq& \frac{1}{(4\pi)^2} \left( \frac{10^2 \; \hbox{GeV}}{ 10^{-3} \; \hbox{eV}} \right)^4 \simeq 10^{54} \,.
\eea
These are clearly still much too large, by enormous factors.\footnote{And because we know the properties of these particles, the values of $C_a$ are absolute predictions, and cannot be adjusted to provide cancellations as might have been done (using supersymmetry, for example, for any hypothetical particles up at the Planck scale.}

But why can't these contributions also be absorbed into a finite (but large) renormalization of the classical contribution? In principle they can, although (as is now argued) this does not provide the relief one might hope.

\subsection{A more effective formulation of the `Old' CC problem}
\label{ssec:MoreEffectiveOld}

A flaw with the previous discussion is that it remains a bit too old school. In particular it talks about {\em the} classical action (and quantum corrections to it) as if Mother Nature gives us a fundamental classical action that we then quantize to make predictions.

But in the modern, more physical, picture of renormalization there is not really a unique `classical' theory that gets quantized at all. Instead, what one usually calls the classical action is really a fully quantum quantity --- the Wilson action --- that is appropriate for describing physics below a given energy scale, $\mu$. In principle, it could be obtained from a more microscopic theory of physics at higher energies by integrating out all of the states having energies larger than $\mu$. In this construction the action, $\cS$, for the microscopic theory is what we normally call the `classical' action, from which the Wilson action is obtained by integrating out heavy degrees of freedom.\footnote{The description here follows that of the review \cite{EFTrev}.}

For example, suppose $h^a$ denotes a collection of quantum fields describing states with energies $E > \mu$, and $\ell^i$ denotes a collection of quantum fields describing the states with energies $E < \mu$. Then the generating functional for correlation functions involving only the light states can be represented in terms of a path integral of the form
\bea \label{WilsonUse}
 \exp \Bigl( i W[J] \Bigr) &=& \int \cD h \cD \ell \; \exp \left( i \cS[\ell, h] + i \int \exd^4x \, J_i \ell^i \right) \nn\\
 &=& \int \cD \ell \; \exp \left( i S_\ssW[\ell] + i \int \exd^4x \, J_i \ell^i \right) \,,
\eea
where
\be \label{WilsonDef}
 \exp \Bigl( i S_\ssW[\ell] \Bigr) := \int \cD h \; \exp \Bigl( i \cS[\ell, h] \Bigr) \,,
\ee
defines the Wilson action, $S_\ssW$, for physics below the scale $\mu$. The second line of eq.~\pref{WilsonUse} shows that it appears in the expression for $e^{iW[J]}$ (and so also for observables involving only the $\ell^i$ states) in precisely the way a classical action normally would. Furthermore, it can be shown \cite{EFTrev} that in many circumstances of interest --- {\em i.e.} when there is a clear hierarchy of scales between the states described by $h^a$ and those described by $\ell^i$ --- $S_\ssW$ can be written as an integral over a local lagrangian density, $S_\ssW = \int \exd^4x \, \cL_\ssW$, built from polynomials of the fields and their derivatives.

Although the coupling constants appearing in $\cL_\ssW$ depend explicitly on $\mu$, this $\mu$-dependence is guaranteed to cancel with the explicit $\mu$-dependence associated with the subsequent integration over the light fields $\ell^i$. Notice that this is reminiscent of what happens with the ultraviolet cutoff, $\Lambda$, in renormalization theory: the explicit $\Lambda$-dependence arising when integrating out fields cancels the $\Lambda$-dependence that the bare couplings are assumed to have when expressed in terms of renormalized couplings.

Furthermore, the process of `integrating out' heavy states like $h^a$ is clearly recursive, and the Wilson action, $\hat S_\ssW$, for another scale $\hat\mu < \mu$ can be obtained from the Wilson action, $S_\ssW$, defined for the scale $\mu$, by integrating out that subset of $\ell^i$ states whose energies lie in the range $\hat\mu < E < \mu$. The new action, $\hat S_\ssW$, has an equally good claim to be the `classical' action for observables involving only energies $E < \hat \mu$. From this point of view, the original microscopic action, $\cS$, may just be the Wilson action defined at the cutoff scale $\Lambda$ in terms of an even more microscopic action that applies at still higher scales.

When computing only low-energy observables there is therefore no fundamental reason to call `the' microscopic action, $\cS$, the classical action instead of the Wilson actions, $S_\ssW$ or $\hat S_\ssW$. All have equally good claims, and only differ in the range of scales over which they can be used to compute physical quantities.

Returning now to the `old' cosmological constant problem, although it is possible to absorb the finite $m$-dependent contributions into a renormalization of the classical action, once this is done once it cannot be done again for the effective cosmological constant in the Wilson action at other scales. For which classical action should we do it?

The real answer is that we shouldn't do it for any of them. What renormalization is telling us is that there is no special classical theory, and so if a physical parameter should happen to be small its smallness should be understandable in {\em any} of the possible effective Wilson actions in which we care to ask the question.
Indeed, this is the way hierarchies of scale usually work. For instance, if we wish to understand why atoms are larger than nuclei we can ask the question in terms of the couplings of the Wilson action appropriate to any scale. If we use the Standard Model, defined at hundreds of GeV, then the size of an atom is set by the Bohr radius, $1/a_0 \simeq \alpha m_e$, and the size of nuclei is set by the QCD scale, $1/r_\ssN \simeq \Lambda_\QCD$. Atoms are larger than nuclei because the fine-structure constant is small, $\alpha \simeq 10^{-2}$, and the electron is much lighter than the QCD scale, $m_e/\Lambda_\QCD \simeq 10^{-3}$.

Now suppose we ask the same question in the effective theory below the confinement scale of QCD, where the quarks and gluons of the Standard Model are replaced by the protons and neutrons (or nucleons) that are built out of them. Although the Bohr radius is still set by $\hat\alpha \hat m_e$ in this new theory, the size of a nucleus is now set by the nucleon mass, $\hat m_\ssN$, where `hats' denote the corresponding renormalized parameters within this new Wilson action. The quantities $\hat \alpha$, $\hat m_e$ and $\hat m_\ssN$ can be computed in terms of the parameters $\alpha$, $m_e$ and $\Lambda_\QCD$, of the Standard Model, and when this is done they still satisfy $\hat \alpha \ll 1$ and $\hat m_e / \hat m_\ssN \ll 1$.

In general, there are two parts to understanding why any particular physical parameter (like the cosmological constant or a scalar mass or some other coupling) might be small. One must ask:
\begin{enumerate}
\item Why is the parameter small in the underlying microscopic theory, like $\cS$?
\item Why does it {\em remain} small as one integrates out the higher energy modes to obtain the Wilson action for the effective theory appropriate to the energies where the parameter is measured?
\end{enumerate}
When both of these questions have an answer then the small parameter is said to be {\em technically natural}. Our understanding of why atoms are large compared with nuclei is technically natural in this sense. Unfortunately, both of these questions do {\em not} have an answer for the cosmological constant within the Standard Model, and so our understanding of the small size of this parameter is not technically natural.\footnote{There is another, slightly more specific, criterion that is sometimes known as `technical naturalness', that is phrased in terms of symmetries. This other criterion is here called 't Hooft naturalness, and is described in more detail in a later section. So in the terminology of this review 't Hooft naturalness is sufficient for technical naturalness, but needn't be equivalent to it.}

In the effective theory well below the milli-eV scale, which is implicitly used in cosmology, the effective cosmological constant can indeed be renormalized to be of order the observed dark energy density: $V_{0\,{\rm le}} \simeq \rho_{\rm vac} \simeq (10^{-2} \; \hbox{eV})^4$. But this then also tells us how big the cosmological constant, $V_{0\,{\rm he}}$, must be in the Wilson action for the effective theory above the electron mass. Because the electron is present in this high-energy theory, but not in the lower energy one at sub-eV energies\footnote{More properly, because the electron is stable it can be present in the low-energy theory by restricting to states having a definite lepton number. But its antiparticle, the positron, is integrated out in this effective theory, precluding there being large quantum corrections to the vacuum energy from electron-positron loops.} $V_{0,{\rm le}}$ and $V_{0,{\rm he}}$ must be related by a formula like eq.~\pref{rhovactheoryR}:
\be \label{rhovactheoryW}
 \rho_{\rm vac} \simeq V_{0,{\rm le}} \simeq V_{0,{\rm he}} + \left( C_{e} m_e^4  + \cdots \right) \,.
\ee
The only way $V_{0,{\rm le}}$ can be small enough is to have $V_{0,{\rm he}}$ and the electron quantum correction cancel one another to an accuracy of better than 30 decimal places! And the required cancellation only gets worse as one moves to the Wilson action defined at energies above the masses of even heavier particles.

We know of no other hierarchies of scale in Nature that work like this, and that is the more precise reason why predictions like eq.~\pref{rhovactheoryR} or eq.~\pref{rhovactheoryW} are really a problem.

Solving this problem is clearly going to be hard; it involves modifying how even the electron contribution to the vacuum energy gravitates, and moreover this must be done at {\em low} energies right down to sub-eV scales. But the electron is probably the particle that we think we understand the best, and this modification must be done in such a way as not to ruin any of the many successful predictions that have been made of electron properties at these energies. This seems a tall order (though it may yet be possible, as argued below).

On the other hand, the upside to the need to modify physics at very low energies is that any successful proposal is likely to be testable in many ways using a variety of low-energy experiments. If there were only at least one solution to the problem, it would be very predictive. In practice it has not yet been possible to profit from this observation in the absence of any convincing candidate solutions (more about which below).

\subsection{Criteria for a successful solution}
\label{ssec:Criteria}

Before descending into a partisan description of some of the approaches towards solving the problem, it is worth having in mind as benchmarks the minimal features that any successful solution must have. To be successful any candidate solution must:

\begin{enumerate}
\item {\em Go beyond the classical approximation.} At the purely classical level there is no problem at all with having a small vacuum energy. Since the data seems perfectly happy with dark energy described by a cosmological constant, and since a cosmological constant is by far the simplest proposal, Occam's razor makes it hard for any purely classical alternative to be regarded as an improvement.
\item {\em Apply at scales larger than an eV.} Since the quantum problem only arises for particles whose mass is larger than sub-eV scales, a candidate theory must be applicable at these scales in order to be able to check that the problem is solved. Put another way, there is no problem at the quantum level with a simple cosmological constant if all particles in the theory have masses below an eV or so. Again Occam's razor argues that a cosmological constant wins when compared with a more complicated theory that applies only below eV scales.
\item {\em Do no harm.} It is not sufficient to have a small quantum vacuum energy if this is done at the expense of contradicting other things we know about Nature. As we shall see, there {\em are} symmetries that can suppress a vacuum energy when unbroken (such as supersymmetry or scale invariance), but these also have other unacceptable predictions when unbroken (degenerate superpartners or vanishing masses) and so do not do what is required. Similarly, a successful proposal should not introduce new naturalness problems, such as small scalar masses, unless these are also stable against large quantum corrections.
\end{enumerate}
As a rule of thumb, a good way to tell whether a candidate theory is viable is to see if it is able to include the electron, and so to allow an understanding of how its contributions to the vacuum energy can be small without destroying other things we know about electrons. This is, after all, what makes solving the problem hard.

\subsection{Other CC problems}
\label{ssec:othercc}

Having established what the problem is, and criteria for a successful solution, the next section briefly examines a few approaches that are well-studied in the literature and then closes with the direction that I personally believe to be the most promising. Before doing so, however, first a brief aside to describe some of the other cosmological constant problems about which people also worry.

The `new' cosmological constant problems have their origin in the discovery that the vacuum energy is not exactly zero. They usually presuppose there is a solution to the `old' cosmological constant problem that makes the vacuum energy precisely zero. Then the thing to be explained is why it is not {\em precisely} zero, and instead has a nonzero but small value.

This `new' problem takes a different form depending on whether or not the nonzero dark energy density is regarded to be time-dependent or time-independent. In the time-independent case the goal is to have a reason why the dark energy density should take a value in the milli-eV regime. The trick is to obtain it by combining the other scales in the problem. This is often done by using the numerology that relates the milli-eV scale to the electroweak scale (which we take to be of order $M_\EW = 1$ TeV) and the Planck mass,
\be
 \frac{M_\EW^2}{M_p} \simeq \frac{\left( 10^3 \; \hbox{GeV} \right)^2}{10^{19} \; \hbox{GeV}} \simeq 10^{-4} \; \hbox{eV} \,.
\ee

The situation is slightly different if the dark energy density is thought to be time-dependent, such as might happen if it is described by the evolution of a scalar field. In this case the dark energy density can vary enormously as the universe expands, much as do the radiation and matter energy densities, making it remarkable that matter and radiation happen to be so very close in size at the present epoch. This observation is often called the `Why Now?' (or `Coincidence') problem. Ideally this would be ensured in a way that does not depend on minutely adjusting initial conditions, perhaps by arranging the scalar dynamics to evolve towards an attractor solution at late times \cite{DErevs}.

Time-dependent dark-energy cosmologies usually also have another quantum-corrections problem, particularly if the time-dependence is associated with an evolving scalar field. This is because having significant scalar evolution over cosmological times usually requires the scalar mass to be much smaller than the Hubble scale during the epoch of scalar evolution: $m \ll H$.

Such a small scalar mass is a problem because quantum corrections to scalar masses also tend to be proportional to the largest scales to which the scalar couples. Indeed, a naive calculation of the quantum correction to a scalar mass due to a quantum correction involving another particle of mass $M$ gives a quadratically divergent result,
\be
 \delta m^2 \simeq \left( \frac{g}{4\pi} \right)^2 ( k_0 \Lambda^2 + k_1 M^2 + \cdots ) \,,
\ee
where $g$ is the coupling strength between the light scalar and the virtual mass-$M$ particle and the $k_i$ are order unity (due to the explicit factor of $1/(4\pi)^2$ out front). As before the $\Lambda^2$ term can be renormalized into the classical scalar mass parameter, leaving the sub-leading terms proportional to the square of the renormalized mass,\footnote{The reasoning as to why this cannot be absorbed into finite renormalizations is the same as given earlier for the cosmological constant.} $M^2$.

To get a feel for how big a mass $M$ would be a problem, imagine we demand $\delta m^2$ be smaller than the present-epoch Hubble scale, $H^2 \simeq (\rho_{\rm vac}/M_p^2) \simeq ( 10^{-34}  \; \hbox{eV})^2$. Suppose also we imagine minimizing the size of corrections by making the coupling to heavier fields extremely weak, such as of order gravitational strength: $g \simeq M/M_p$. Then $m \simeq g M/4\pi \simeq M^2/(4\pi M_p) < H$ implies
\be
 M < \sqrt{4\pi H M_p} \simeq \sqrt{4 \pi} \; \rho_{\rm vac}^{1/4} \simeq 10^{-2} \; \hbox{eV} \,.
\ee
Since this is much smaller than any particles we know about, it seems as if even gravitational-strength couplings to ordinary particles should generically generate too-large contributions to scalar masses.\footnote{A key word here is `generically', and there are exceptions to these estimates \cite{naturallylight}, so the size of these corrections must be evaluated on a case-by-case basis for any candidate theories.}

\section{Possible approaches}
\label{section:approaches}

This section briefly describes some of the proposals that have been made over the years, with a view to comparing how they stand up compared with the three benchmarks given at the end of the previous section. The purpose here is not to be exhaustive, given the enormous number of models that have been proposed over the years, and given the exemplary job describing these by other lectures at this school. In particular, the focus here is on those that are aimed specifically at the `old' cosmological constant problem, rather than the `new' ones. The section closes with a brief cartoon of the direction which I believe to be the most promising approach to date.

\subsection{Roads well travelled}
\label{ssec:roadstrav}

We start with some of the approaches that have been relatively well explored over the years.

\subsubsection*{Head in the sand}

Probably the most well-established attitude to the cosmological constant problem in particle physics is one of pragmatic despair. Since there are no candidate solutions to choose amongst, one soldiers on with other problems in physics and hopes that whatever solves the cosmological constant problem is not going to be important to this other physics.

This is a harder point of view to adopt in cosmology, where a commitment must be made as to how or whether the dark energy clusters and/or evolves with time.

\subsubsection*{Anthropic arguments}

A more refined point of view building on the absence of known solutions is to take this absence as evidence that quantum corrections to the vacuum energy need not be small after all \cite{WbgNoGo, Anthropic}. That is, one denies that both of the questions given in \S\ref{ssec:MoreEffectiveOld} must be answered for the cosmological constant, and simply accepts that there is a very precise cancellation that occurs between the renormalized classical parameter, $V_0$, and the quantum contributions to $\rho_{\rm vac}$. As emphasized earlier, this is a logically consistent point of view, though it is radical in the sense that it is the first known example where this occurs for parameters in the Wilson action.\footnote{There are examples of coincidences of scale that do not require a fundamental explanation, such as the apparent sizes of the Sun and the Moon as seen from the Earth. However I do not know of any examples of this type that involve the smallness of parameters in a Wilson action.}

There is a better face that can be put on this cancellation if the microscopic theory has three features. First, the microscopic theory could have an enormous number of candidate vacua, with the cosmological constant differing from vacuum to vacuum. (This is actually likely to be the situation at a fundamental level, if string theory is any guide.) Second, the microscopic theory might have a reason to have sampled many of these vacua somewhere in space at some time over the history of the universe. (This is also not far-fetched in theories that allow eternal cosmic inflation within a complicated potential energy landscape, such as seems likely for string theory.) Third, it might be true that observers can only exist for a very small range of vacuum energies, not much different from the observed dark energy density.

In this case one might expect the universe to be populated with an enormous number of independent regions, in each of which a particular vacuum (and cosmological constant) is selected. The vast majority of these vacua do not have observers within them whose story needs telling, but those that do can only have a small cosmological constant since this is required for the observers to exist in the first place. Since we live in such a world we should not be surprised to find evidence for dark energy in the range observed.

Although this may well be how things work, most (though not all) of its proponents would prefer to have a technically natural solution to the problem (satisfying the two criteria of \S\ref{ssec:MoreEffectiveOld}) if only this were to exist. There are two dangers to adopting this kind of anthropic approach. One is that it becomes a dogma that stops people searching for a more traditional solution to the problem. Another is that it is difficult to know how to falsify it, and what the precise rules are that one should use when making predictions. (Of course this is partly the point: it is not clear how one makes predictions in theories having an enormous landscape of possible vacua, and it is important that this gets thought through to see if a sensible formulation can be found.)

My own view on this is to accept that there is an important issue to be resolved to do with making predictions in theories (like string theory) that have a complicated landscape. But in practice the problem right now is not that we know of too many acceptable vacua, it is that it is hard to find any good vacua at all given the large number that must be sorted through. (Once we have two examples that include the Standard Model we can start worrying about their statistics.) One thing that might help in this search is to have `modules' that build in features we know to be true of the world around us. These modules include the Standard Model particle content, some candidate for dark matter, and hopefully could include a natural description of dark energy if this could be found.

\subsubsection*{Symmetries?}

The traditional approach to satisfying the two technical-naturalness criteria of \S\ref{ssec:MoreEffectiveOld} is to identify a symmetry or an approximate symmetry that enforces the smallness of the parameter of interest. This happens in particular if the theory would have more symmetry when the small parameter actually vanishes than it does when it doesn't vanish --- a criterion that can be called {\em 't Hooft naturalness} after its original proposer \cite{tHooft}. When this is true the enhanced symmetry prevents quantum corrections from generating a nonzero parameter if it is initially taken to be zero. This typically ensures that corrections to an initially small parameter must then themselves also be small.

With this in mind one might ask: are there any symmetries that can forbid a nonzero cosmological constant? It turns out there are at least two, though none has yet led to a simple solution to the problem.

\medskip\noindent
{\em Supersymmetry:}

\medskip\noindent
In a footnote below eq.~\pref{rhovactheory} it was mentioned in passing that loop corrections to the vacuum energy can be made to cancel between bosons and fermions, since these contribute with opposite signs. Supersymmetry is the symmetry that enforces this cancellation. Supersymmetry relates particles whose spins differ by $\frac12$, and so (because of the spin-statistics theorem) necessarily relates bosons with fermions. When it is not spontaneously broken\footnote{That is, if the vacuum is invariant under supersymmetry.} the particles related in this way must share the same mass and interaction strengths, and this is why their contributions to the vacuum energy can precisely cancel.

The problem with using this as a solution to the cosmological constant problem is this same fact that the bosons and fermions must have equal masses and interactions in order to precisely cancel, and we do not find in Nature bosons and fermions organized into these kinds of bose-fermi pairs. Observations can only be consistent with a supersymmetric world if supersymmetry is spontaneously broken, since in this case putative bose-fermi partners become split in mass by the supersymmetry-breaking scale, $F_s$:
\be
 m^2_\ssB - m^2_\ssF \simeq g_s^2 F_s^2 \,,
\ee
where $g_s$ is a measure of the strength with which the bose-fermi pair of interest couples to the sector of the theory that spontaneously breaks supersymmetry. The failure to find superpartners for the known Standard Model particles \cite{PDG, SUSYbounds} implies $g_s F_s$ for them cannot be less than of order 1 TeV.

But such a mass splitting also ruins the cancellation of the corresponding vacuum energies, leading to nonzero results of order
\be
 \delta \rho_{\rm vac} \simeq \cC \, (m_\ssB^4 - m_\ssF^4) \simeq \cC  (m_\ssB^2 + m_\ssF^2) g_s^2 F_s^2 \,.
\ee
Because $g_s F_s$ cannot be lower than the TeV scale for Standard Model particles this does not in itself help cancel their contributions to $\rho_{\rm vac}$. But it {\em can} help understand why the vacuum energy is not higher than TeV scales (for instance why it is not of order the Planck scale). So this is an example which satisfies criteria 1 and 2 of \S\ref{ssec:Criteria}, but fails criterion 3.

\medskip\noindent
{\em Scale Invariance:}

\medskip\noindent
Scale invariance is a symmetry where $g_{\mu\nu} \to \zeta^2 \, g_{\mu\nu}$ with constant $\zeta$, in which case the Ricci tensor, $R_{\mu\nu}$, remains invariant while the Ricci scalar transforms as $R = g^{\mu\nu} R_{\mu\nu} \to \zeta^{-2} R$. Consequently the Einstein tensor, $R_{\mu\nu} - \frac12 \, R \, g_{\mu\nu}$, remains invariant, and so transforms differently than does a cosmological constant, $\lambda g_{\mu\nu}$. As a result scale invariance can forbid the vacuum energy from appearing in Einstein's equations.

So far so good. But there are several problems with using this observations to solve the cosmological constant problem. The first of these is the observation that scale invariance is usually not preserved by quantum corrections, even if it is arranged to be a classical symmetry \cite{ScaleAnomaly}. But there is also another problem that arises once this symmetry is extended to matter, even if it were arranged to be preserved at the quantum level.

To see what this second problem is, imagine coupling gravity to a scalar field, $h$, with action as given in eq.~\pref{ScalarS}, including the scalar potential eq.~\pref{ScalarV}. To extend the scale invariance to the scalar sector we must demand $\partial_\mu h$ does not transform when $g_{\mu\nu} \to \zeta^2 g_{\mu\nu}$. To see this notice that this ensures that the derivative terms in the scalar stress energy
\be
 T_{\mu\nu} = \partial_\mu h \, \partial_\nu h - \frac12 \Bigl[g^{\lambda\rho} \partial_\lambda h \partial_\rho h + V(h) \Bigr] g_{\mu\nu} \,,
\ee
are also invariant. This also shows that $V$ must vanish if $h$ is invariant, and although this is consistent with the vanishing vacuum energy it also forces the scalar to be massless. We are not too surprised to find no cosmological constant problem when only using massless particles, and it is because we know that Standard Model particles have masses that we have a problem in the first place (at face value we again run up against criterion 3 of \S\ref{ssec:Criteria}).

But it is actually possible to have nonzero masses if we spontaneously break scale invariance. This can be done by allowing $h$ to transform, according to $h \to h + f \ln \zeta$, for a constant mass scale $f$, which is allowed since it leaves $\partial_\mu h$ invariant. Once this is done, particles can be given masses in a scale invariant way by coupling them appropriately to $h$.

Consider, for example, adding a second scalar field, $\phi$, which does not transform under scale transformations. In this case the stress energy for both $h$ and $\phi$ becomes
\be
 T_{\mu\nu} = \partial_\mu h \, \partial_\nu h + \partial_\mu \phi \, \partial_\nu \phi - \frac12 \Bigl[g^{\lambda\rho} \partial_\lambda h \partial_\rho h + g^{\lambda\rho} \partial_\lambda h \partial_\rho h + V(h, \phi) \Bigr] g_{\mu\nu} \,,
\ee
and the scalar potential can also be invariant provided it depends on $h$ in the specific way dictated by scale invariance:
\be
 V(h, \phi) = U(\phi) \, e^{-2 h/f} \,.
\ee
We can give $\phi$ a mass if we take $U(\phi) = \frac12 \, \mu^2 \phi^2$, since in this case the potential is minimized at $\phi = 0$ and $h = h_0$ for any $h_0$. Evaluating the second derivatives of the potential at this minimum shows that $\phi$ has mass $m = \mu \, e^{-h_0/f}$ while $h$ remains exactly massless.

What looks more promising for the cosmological constant is that scale invariance insures that $V$ itself vanishes at this same minimum, regardless of the value of $h_0$ taken by $h$ in the vacuum, and in a way that is consistent with the existence of a mass for the field $\phi$. These can coexist because spontaneous scale breaking allows masses to be nonzero, but {\em continues} to enforce the vanishing of the potential at its minimum. The reason it does so is that both masses and $V$ had to vanish as $h_0 \to \infty$, since this corresponds to the case where scale invariance is unbroken (so {\em all} dimensionful quantities vanish in this limit). But this ensures that $V$ must vanish along the entire one-parameter set of minima of $V(h,\phi)$. All of these vacua must have precisely the same energy because they are all related to one another by a symmetry, scale invariance itself.

But there does remain a fly in the ointment, as is most famously formulated as Weinberg's `no-go' result \cite{WbgNoGo}. The source of the problem can be seen by considering the most general possible form for $U(\phi)$, such as would be expected to be generated once quantum corrections are included. Then, for small $\phi$, we should really consider $U(\phi) = U_0 + \frac12 \, \mu^2 \phi^2 + \frac14 \, \lambda \phi^4 + \cdots$ and so on. In particular, once the vacuum, $\phi = \phi_0$, is found by minimizing $U$, $U'(\phi_0) = 0$, the value $U_{\rm vac} = U(\phi_0)$ is generically nonzero. But this then lifts the flat direction for $h$, which must minimize the potential $V(h) = U_{\rm vac} e^{-2h/k}$, and so is driven to the scale-invariant point, $h \to \infty$.

The good news is that scale invariance always ensures a vanishing vacuum energy, even if spontaneously broken. But because we know particles exist with nonzero masses the scale invariance must be spontaneously broken. This means the scalar potential must have a one-parameter family of minima, related by scale invariance, for all of which the potential vanishes. But the bad news is that quantum corrections --- {\em even if these are scale invariant} --- generically lift this flat direction, leaving only the single minimum at the scale-invariant point (where all masses vanish). The cosmological constant problem, for these theories, is the question of why this lifting does not happen once quantum effects are included.

\subsubsection*{Modifying Gravity?}

Since the evidence for dark energy comes entirely from its gravitational effects, and these are inferred assuming the validity of General Relativity, it is almost irresistible to try to modify gravity to obviate the need for dark energy.

There are a great many models of this type, as was well described by other lectures in this school. Most of these aim at modifying the gravitational field equations in such a way as to allow accelerating cosmological solutions without requiring a specific (matter or vacuum) dark energy density as a source. These models are aimed at the `new' cosmological constant problems, inasmuch as they try to explain the size of the observed acceleration under the assumption that there is not also a cosmological constant. Such models usually do not explain why the cosmological constant should be absent in the first place.

There is an exception to this statement, however. Some modifications to gravity are proposed with the goal of having gravity `screen' the cosmological constant by allowing near-flat cosmologies to coexist with a large vacuum energy. Among the best-motivated examples of this type are those which propose that the graviton might have a very small mass \cite{GMassCC}. This mass is usually taken of order the present-day Hubble scale, $m_g \simeq H_0 \simeq 10^{-33}$ eV, so that it is relevant on cosmological scales but too small to affect tests of gravity over smaller distances, where it is better tested.

Superconductors provide a motivation for hoping a graviton mass might help screen a cosmological constant. Because photons are massless any nonzero current density gives rise to either electric or magnetic fields, because of the Maxwell equation
$\nabla_\mu F^{\mu\nu} = J^\nu$. But in a superconductor the photon acquires a mass because the ground state spontaneously breaks the (gauge) symmetry that makes it massless in the vacuum. Such a mass term modifies the Maxwell equation in a superconductor to become
\be
 \nabla_\mu F^{\mu\nu} - m_\gamma^2 A^\nu = J^\nu \,,
\ee
where $A_\mu$ is the gauge potential from which the field strength $F_{\mu\nu} = \partial_\mu A_\nu - \partial_\nu A_\mu$ is defined.\footnote{The potential can appear because of the spontaneous breaking of gauge invariance in the superconductor.} In this case the presence of a current need {\em not} imply the presence of electric or magnetic fields, because it can instead produce nonzero $A_\mu$ in a way that doesn't give nonzero $F_{\mu\nu}$. It is this phenomenon that underlies the well-known expulsion of magnetic fields from superconductors.

The idea is similar for gravity, with the hope that a graviton mass can remove the requirement that a cosmological constant must generate large curvatures. Linearized about flat space, $g_{\mu\nu} = \eta_{\mu\nu} + 2\kappa h_{\mu\nu}$, a graviton mass is expected to modify the Einstein field equations from $\Box h_{\mu\nu} = \kappa T_{\mu\nu}$ (in an appropriate gauge) to
\be
 \Box h_{\mu\nu} - m_g^2 h_{\mu\nu} = \kappa T_{\mu\nu} \,.
\ee
If so, the choice $T_{\mu\nu} = - \rho_{\rm vac} \eta_{\mu\nu}$ could just imply $h_{\mu\nu} = (\rho_{\rm vac}/m_g^2) \eta_{\mu \nu}$ instead of driving a nonzero curvature: $R_{\mu\nu} - \frac12 \, R \, g_{\mu\nu} \propto \Box h_{\mu\nu} \simeq 0$.

It is yet to be seen whether this mechanism can be fully realized at the nonlinear level, and although efforts are ongoing no proposals have yet succeeded in doing so while satisfying the three criteria of \S\ref{ssec:Criteria}. There are several kinds of problems encountered. On one hand, if the physics that gives the graviton a mass is weakly coupled, then the three additional spin states required to obtain a massive spin-2 field from the two helicity states of the massless graviton survive to mediate long-range interactions that are ruled out by solar-system tests of gravity. On the other hand, if these states are expunged --- as can be done by having them strongly couple \cite{Vainshtein} --- the resulting effective description appears to break down (requiring some sort of unknown ultraviolet completion) at energy scales that are much lower than the eV scales of the cosmological constant.

Although the final word is not in on these proposals, at our present state of knowledge they fail to satisfy criteria 2 and/or 3 of \S\ref{ssec:Criteria}, depending on the formulation.

\subsection{A way forward?}
\label{ssec:forward}

So is it hopeless? What direction, if any, has the best prospects of providing a technically natural cosmological constant? I believe there is one line of inquiry that is very promising and has not yet been explored to its end. My collaborators and I have been exploring this direction, with encouraging results so far, though the final verdict is not yet in. This section is devoted to describing the main ideas.

Recall what is the essence of the problem: we believe quantum effects generate large vacuum energies, and the vacuum's Lorentz invariance automatically gives this the $w=-1$ equation of state of a cosmological constant: $T_{\mu\nu} = - \rho_{\rm vac} g_{\mu\nu}$. But cosmologists measure very small curvatures when they detect the acceleration of the universe's expansion. The conundrum is that these are directly equated in Einstein's equations, eqs.~\pref{EinsteinCC} or \pref{RicciCC}.

We seek something that can break this direct link between the vacuum energy and the curvatures measured in cosmology. Moreover, we must do so only for very slow processes (like the vacuum) and not also for fast ones (like atoms) \cite{Bandpass}. Fast quantum fluctuations should gravitate in an unsuppressed way because we know that such fluctuations do contribute to atomic energy levels in atoms. They do so for many reasons, ranging from the (order-unity) contribution to the nucleon mass coming of the quark and gluon zero-point energies, down to the small (but not insignificant) Lamb shift coming from electron fluctuations. For Hydrogen-like atoms the Lamb shift contribution is of order \cite{LambShift}
\be
 \delta E \simeq \frac{\alpha (Z\alpha)^4}{4\pi n^3} \, F(Z\alpha) \, m_e \,,
\ee
to the total mass of an atom, where $\alpha$ is the fine-structure constant, $Z$ is the atom's nuclear charge, $n$ is the principal quantum number of the energy level of interest and $F(Z\alpha)$ is a slowly-varying function that ranges from order unity when $Z =1$ to about 400 when $Z=100$. This represents a fraction of the atom's rest mass that is of order $4 \times 10^{-14}$ for Beryllium and $9 \times 10^{-12}$ for Titanium. But tests of the principle of equivalence, such as those using lunar laser ranging \cite{LLR, Will}, show that gravity couples to the {\em total} energy density of an atom, including its binding energy, to an accuracy of a part in $10^{13}$, and so could in principle tell if, for some reason, quantum fluctuations within atoms did not gravitate.

\subsubsection*{The extra-dimensional loophole}

The good news is there is a loophole within which it is possible to break the link between vacuum energy and curvature, and so can seek a solution. This loophole is based on the observation that Lorentz invariance plays an important role in formulating the problem, because it so severely restricts the form of the vacuum stress energy (at least in 4 dimensions). But the situation would be different if there were extra dimensions because then we would only know that the vacuum must be Lorentz invariant in the four dimensions that we can see. We also would only really know that the curvatures must be small in these same dimensions (which are the ones we can access in cosmology). And although it remains true that the vacuum stress energy must curve something, in extra-dimensional models nothing says it must curve the dimensions we see.

The gravitational field of a cosmic string in four dimensions provides the simplest example that illustrates this loophole more explicitly. Imagine that a relativistic cosmic string sweeps out a world-sheet along the $z-t$ plane, transverse to the $x-y$ directions. The stress energy of such a string is Lorentz invariant in the $z-t$ directions, $T_{ab} = - \cT g_{ab} \, \delta^2(x)$, where $\cT$ is the string's tension ({\em i.e.} mass per unit length), and $a,b$ denotes the $z-t$ directions along the string world-sheet. The gravitational field sourced by this stress energy is known \cite{CosmicString} and the geometry of the two transverse directions turns out to be a cone, whose apex is located at the position of the string. The tension on the string gives rise to a curvature singularity there, $R \propto \kappa^2 \cT \, \delta^2(x)$, appropriate to a cone. But what is important is that the geometry along the two Lorentz-invariant directions remains perfectly flat, regardless of the precise value of $\cT$.

This suggests trying similar examples having two more dimensions (six dimensions in total) with the 2-dimensional string world-sheet being replaced by the world-volume of a 4-dimensional Lorentz-invariant brane, situated at specific points within two compact extra dimensions. In the simplest examples the two extra dimensions have the geometry of a sphere, and there is a brane located at both the sphere's north and south poles (at which points the transverse curvature also has conical singularities, like for a cosmic string). All standard model particles are imagined to be confined to one of these branes, whose tension ({\em i.e.} vacuum energy per unit volume) is not particularly small --- of order $(1 \; \hbox{TeV})^4$. The hope is that the geometry seen by an observer on the brane (us) can remain flat regardless of the size of the brane vacuum energy density.

\medskip\noindent
{\em A false start: Extra dimensions alone}

\medskip\noindent
In fact, explicit solutions to higher-dimensional field equations of this type do exist, the simplest arising as solutions to the Einstein-Maxwell system in six dimensions. Although this is not quite yet what is needed, it is useful to describe it to set up what is the real proposal. The simplest action for Einstein-Maxwell theory coupled to two source branes is
\be
 S = - \int \exd^6x \sqrt{-g} \; \left( \frac{1}{2\kappa_6^2} \; \cR + \frac14 \, F_{\ssM\ssN} F^{\ssM \ssN} + \Lambda_6 \right)
 - \sum_{b=N,S} \int \exd^4 x \sqrt{-\gamma} \; \cT_b \,,
\ee
where $\cR$ is the 6D curvature scalar, $\kappa^2_6 = 8 \pi G_6$ is the 6D gravitational constant and $\Lambda_6$ is a higher-dimensional cosmological constant. $\gamma_{ab} = g_{\ssM \ssN} \partial_a x^\ssM \partial_b x^\ssN$ is the induced metric on the branes, and $\cT_b$ denotes the tension on the brane labelled by the binary label, `$b$', (which will correspond to label at which pole of the spherical extra dimensions the corresponding brane sits).

Solutions to the resulting field equations can be sought of the form
\be \label{FreundRubin}
 \exd s^2 = e^{2W(y)} \hat g_{\mu\nu}(x) \exd x^\mu \exd x^\nu + g_{mn}(y) \exd y^m \exd y^n
 \quad \hbox{and} \quad
 F_{mn} = f \epsilon_{mn} \,,
\ee
where $W$ is the `warp factor' and $\epsilon_{mn}$ denotes the antisymmetric Levi-Civita tensor constructed using the metric $g_{mn}$. The indices $\mu, \nu$ run over the four on-brane dimensions, $m,n$ run over the two transverse dimensions and $M,N$ run over all six dimensions. The Maxwell field is present to stabilize the extra dimensions, since gravity likes the extra dimensions to shrink. But flux quantization for the Maxwell field implies that its integral over the two transverse dimensions is an integer
\be \label{FluxQ1}
 \int_{M_2} F = \int \exd^2 x \sqrt{g_2} \; f = \frac{2\pi n}{e_6} \,,
\ee
where $e_6$ is the Maxwell field's coupling constant and $g_2 = \det \, g_{mn}$. Shrinking the extra dimensional space, $M_2$, requires $F_{mn}$ to grow, and it is the energy cost of this growth that competes with gravity to fix the extra-dimensional size.

To see how this works in detail, suppose (for simplicity) that both the `north' and `south' branes share the same tension, $\cT$. Then the solutions to the field equations are particularly simple: spheres of the form \cite{NonSUSYRugby}
\be \label{2Dmetric}
 g_{mn} \exd x^m \exd x^n = L^2 \Bigl( \exd \theta^2 + \beta^2 \sin^2 \theta \, \exd \xi^2 \Bigr) \quad \hbox{and} \quad
 F_{\theta\xi} = Q \beta L^2 \sin \theta \,,
\ee
where $L$ is the extra-dimensional radius and $1-\beta = \kappa_6^2 \cT/2\pi$ measures the size of the conical defect at the north and south pole.\footnote{To see why $\beta$ encodes the conical defect absorb $\xi \to \beta \xi$ to remove $\beta$ from eq.~\pref{2Dmetric}, at the expense of making $\xi$ periodic with period $2\pi \beta$. The conical defect converts the sphere into something more the shape of a rugby ball (or of an American football).} The metric in the on-brane directions can be taken to be maximally symmetric ({\em i.e.} 4D de Sitter, Minkowski or anti-de Sitter geometries), and so its entire (Riemann) curvature is completely dictated by its Ricci scalar, $\hat R$.

The field equations and flux quantization then give $W = 0$ while $L$, $Q$ and the 4D curvature scalar, $\hat R$, are given in terms of $\beta$, as follows
\be \label{NonSUSYsoln}
 Q = \frac{n}{2 \beta e_6 L^2} \,, \quad \hat R = \kappa_6^2 (Q^2 - 2 \Lambda_6) \quad \hbox{and} \quad
 \frac{1}{L^2} = \frac{8 \beta^2 e_6^2}{3n^2 \kappa_6^2} \left[ 1 \mp \sqrt{1 - \frac{3 n^2 \kappa_6^4 \Lambda_6}{8 \beta^2 e_6^2}} \right] \,.
\ee

Clearly the on-brane directions are in general curved, but can be made flat using a specific choice for $\Lambda_6$. After doing so, we can imagine tweaking the brane tensions, $\cT \to \cT + \delta \cT$ and asking how the 4D curvature changes. Since $\cT$ enters the solutions only through $\beta$, we shift $\beta \to \beta + \delta \beta$ (with $\delta \beta = -\kappa_6^2 \delta \cT/2\pi$) and linearize the formulae \pref{NonSUSYsoln} in $\delta \beta$ (without re-tuning $\Lambda_6$ to keep $\hat g_{\mu\nu}$ flat).

When this is done we find no joy: the on-brane curvature turns out to be
\be
 \hat R = 2 \kappa_6^2 Q \delta Q =  \frac{4 \delta \beta}{\beta L^2} = - \frac{2 \kappa_6^2 \delta \cT}{\pi \beta L^2}  \,.
\ee
Physically, this curvature has the following origin. Changing $\beta$ changes the volume of the extra dimensions because it changes the conical defect angle. But flux quantization then requires the Maxwell field to change to compensate for the change in volume. It is the energy of this flux change, $\delta Q$, that ultimately causes $\hat R$ to become nonzero, and it is comparatively large because the homogeneous flux configuration makes the energy cost proportional to the extra-dimensional volume.

How large is this curvature? Using the relation between the 4D and 6D gravitational coupling, $\kappa_6^{2} = 4 \pi \beta L^2 \kappa^{2}$, we may write $\hat R = - 8 \kappa^2 \delta \cT$, and when compared with eq.~\pref{RicciCC} this shows that we have a 4D curvature that is what would have been obtained in 4D from a vacuum energy density $\rho_{\rm vac} = 2 \delta \cT$. This is precisely the size we would have expected in 4D given a change, $\delta \cT$, of vacuum energy on each of two branes; there is no particular extra-dimensional magic here.

\subsubsection*{Doubling Down: Supersymmetric extra dimensions}

The presence of the 6D cosmological constant should have tipped us off that we might not have made much progress on the cosmological constant problem simply by moving to extra dimensions. There is a better chance if the extra-dimensional physics is supersymmetric, however, because in six (and higher) dimensions supersymmetry forbids a cosmological constant (much as would more than one supersymmetry in four dimensions). Because the extra dimensions will turn out to be quite large (of order the final vacuum energy, $1/L \simeq 10^{-2}$ eV, another reason for liking six dimensions is the phenomenological point that only two dimensions can be this large without the higher dimensional Newton's constant being so large the extra dimensions would have already been detected \cite{ADD}. The cosmological constant is asking us to double down; to postulate extra dimensions that are both very large {\em and} supersymmetric ({\em i.e.} Supersymmetric Large Extra Dimensions, or SLED for short \cite{SLED, SLEDrevs, TNCC}).

Notice that we do {\em not} also require the physics on the brane to be supersymmetric, and for almost all purposes we may simply choose only the Standard Model to live on the brane.\footnote{`Almost' here depends on whether or not a dark matter candidate is ultimately chosen to live on the brane too.} Such a brane can nonetheless be coupled consistently to supergravity using the `St\"uckelberg trick'; that is, regarding the non-supersymmetric brane to be supersymmetric, but nonlinearly realized, by coupling a Goldstone fermion --- the Goldstino --- in the appropriate way \cite{SUSYNLR, SUSYBLF}.

This leads to a novel kind of supersymmetric phenomenology \cite{MSLED, SLEDpheno}: a very supersymmetric gravity (or extra-dimensional, bulk) sector (whose supersymmetry breaking scale is of order $1/L \simeq 10^{-2}$ eV) coupled to a particle (brane) sector that is not supersymmetric at all. In particular, the nonlinear realization of supersymmetry on the brane implies that a supersymmetry transformation of a brane particle like the electron gives the electron plus a Goldstino (or, equivalently, a gravitino) rather than a selectron. One does not expect to find a spectrum of MSSM superpartners for the Standard Model, despite the very supersymmetric gravity sector.\footnote{This particular prediction was made \cite{MSLED} before the LHC results showed it to be a huge success.}

There are a variety of supergravities from which to choose in six dimensions, but one choice is particularly convenient in that it allows a rugby ball solution \cite{SS, SLED}, very similar to the one described above. This supergravity is the Nishino-Sezgin chiral gauged supergravity \cite{NS}, whose rugby-ball solutions are described by the following 6D bosonic fields: the metric, $g_{\ssM\ssN}$, a scalar dilaton, $\phi$, and a specific $U(1)_\ssR$ gauge potential, $A_\ssM$. To lowest orders in the derivative expansions, the (supersymmetric) bulk and (non-supersymmetric) brane contributions to the action are
\bea \label{SSUSY}
 S &=& - \int \exd^6x \sqrt{-g} \; \left[ \frac{1}{2\kappa_6^2} \; g^{\ssM\ssN} \left( \cR_{\ssM\ssN} + \partial_\ssM \phi \, \partial_\ssN \phi \right) + \frac14 \, e^{-\phi} F_{\ssM\ssN} F^{\ssM \ssN} + \frac{2 e_6^2}{\kappa_6^4} \; e^\phi \right] \nn\\
 && \qquad\qquad - \sum_{b=N,S} \int \exd^4 x \sqrt{-\gamma} \; \left( \cT_b - \frac12 \, \cA_b \, \epsilon^{mn} F_{mn} + \cdots \right) \,.
\eea
Here $\cT_b$ denotes the brane tension as usual, while $\cA_b$ can be interpreted as describing the amount of Maxwell flux, $\Phi_b = \cA_b \, e^\phi/2\pi$, that is localized on the brane \cite{LargeDimsCurvature}.

The simplest situation is when the two branes are identical, in which case there is a rugby ball solution to these field equations \cite{SS, SLED} of the form of eqs.~\pref{FreundRubin}, but with $W = 0$ and \pref{2Dmetric} replaced by
\be \label{2DmetricSS}
 g_{mn} \exd x^m \exd x^n = \ell^2 \Bigl( \exd \theta^2 + \beta^2 \sin^2 \theta \, \exd \xi^2 \Bigr) e^{-\phi_0} \quad \hbox{and} \quad
 F_{\theta\xi} = Q \beta \ell^2 e^{-\phi_0} \sin \theta \,,
\ee
where $\phi = \phi_0$ is an arbitrary constant, while
\be \label{SUSYsoln}
 \frac{1}{\ell^2} = \kappa_6^2 Q^2 = \left( \frac{2 e_6}{\kappa_6} \right)^2  \quad \hbox{and} \quad \hat R = 0
 \,,
\ee
Several features of this solution are noteworthy:
\begin{enumerate}
\item Inspection of \pref{2DmetricSS} shows that the physical radius of the extra dimensions is now $L = \ell \, e^{-\phi_0/2}$, and so eqs.~\pref{SUSYsoln} imply
\be \label{Lvsphi}
 L^2 e^{\phi_0} = \ell^2 = \left( \frac{\kappa_6}{2 e_6} \right)^2 \,,
\ee
is fixed.
\item Notice that the value of $\phi_0$ is not determined by any of the field equations. This `flat direction' is a consequence of a classical scale invariance of extra-dimensional supergravity, under which $g_{\ssM \ssN} \to \zeta^2 \, g_{\ssM\ssN}$ and $\phi \to \phi - 2\ln \zeta$. (This is why $\phi$ is called the 6D dilaton.) This symmetry is preserved by the branes only if $\cT_b$ is $\phi$-independent and $\cA_b \propto e^{-\phi}$, and is broken otherwise. When broken, the brane interactions lift the flat direction and give a mass to the corresponding KK mode. In the case of most interest for the cosmological constant both $\cT_b$ and $\cA_b$ are $\phi$-independent, and so the breaking of the scaling symmetry by the $\cA_b$ term allows $\phi_0$ to be fixed (in practice, through the appearance of $\cA_b \, e^{\phi_0}$ in the flux-quantization condition --- see below).
\item The presence of localized flux on the brane ({\em i.e.} the $\cA_b$ term) changes the flux quantization condition into
    \be \label{FluxQ2}
     \int_{M_2} F + \frac{1}{2\pi} \sum_b \cA_b \, e^{\phi} = \pm \frac{1}{e_6} \,,
    \ee
    and so
    \be \label{FluxQ2a}
    \Phi_{\rm tot} := \frac{1}{2\pi} \, e^{\phi_0} \sum_b \cA_b = \pm (1 - \beta) \,.
    \ee
    It is this condition that determines the value of $\phi_0$. (Notice that this equation would not have any solutions at all if the $\cA_b$'s were all assumed to be zero, as was often done in early studies of this system \cite{GP}.)
\item Most remarkably, the brane action is flat ($\hat R = 0$) for {\em any} choice of brane lagrangian,\footnote{Explicit solutions where the branes are not identical are also known \cite{DiffBranes, 6Dflatnophi}, and also have flat branes.} and in particular regardless of the value of $\cT_b$. How did this happen?

    The simplest way to see what happens is first to ask why the curvature is flat in the spherical solution in the absence of branes ({\em i.e.} with $\beta = 1$) \cite{SS}. The 4D scalar potential for the fields $\phi_0$ and $L$ is obtained by evaluating the action using the 2D scalar curvature $R = 2/L^2$ and the Maxwell field is $F_{mn} F^{mn} \propto n^2/L^4$. Combining the Einstein and Maxwell actions with the scalar potential\footnote{After first transforming to the 4D Einstein frame: $g_{\mu\nu} \to (1/L^2) g_{\mu\nu}$.} then gives a perfect square in the case with unit flux, $n = \pm 1$:
    \be
     V(L, \phi_0) \propto \int \exd^2 x \sqrt{-g} \; \left( \frac{1}{2\kappa_6^2} \, R + \frac14 \, e^{-\phi_0} F_{mn} F^{mn} + \frac{2 e_6^2}{\kappa_6^4} \, e^{\phi_0} \right) \propto \frac{e^{\phi_0}}{L^2} \left( 1 - \frac{\kappa_6^2}{4 e_6^2 L^2 e^{\phi_0}} \right)^2 \,,
    \ee
    which is minimized at a fixed value of $L^2 e^{\phi_0}$ along which there is a flat direction with $V = 0$, with $e^{\phi_0}/L^2$ not determined.

    Now Einstein's equations imply that adding the brane back-reaction adds a conical singularity to the 2D curvature, $\sqrt{g_2} \; R_{\rm sing} = 2 \kappa_6^2 \sum_b \cT_b \; \delta^2(x-x_b)$, where $\cT_b$ is the brane tension. Using this in the Einstein action (and performing the integral $\exd^2 x$ using the delta function) then gives a contribution of the form $- (1/2\kappa_6^2) \int \exd^2x \sqrt{g_2}\; R = -\sum_b \cT_b$, which precisely cancels the direct contribution of the brane tensions themselves.

    The lesson from this story is that back-reaction is crucial to the process: this cancellation can never be seen working purely within a `probe' approximation where the brane does not perturb its environment.\footnote{A precursor to this phenomenon also occurs in 5D models \cite{5Dversion}, where back-reaction can also be computed. In this case closer inspection \cite{Anti5D} showed that the flat solution arises due to a cancelation with branes whose presence was not explicit but were required to interpret singularities that were necessary on topological grounds. The analogous argument in 6D expresses the extra-dimensional Euler number as the sum of brane tensions plus an integral over extra dimensional curvature. For the rugby-ball geometries of interest here (with the topology of a sphere) this is equivalent to the relation between defect angle and tension used in the text. The situation in 5D is more similar to toroidal compactifications in 6D, where the Euler number condition states that the sum of brane tensions must vanish. Nonetheless, in 6D this is not in itself an obstruction to technical naturalness even for tori. If a tension is changed in a toroidal compactification, the extra dimensions simply curve to satisfy the topological constraint \cite{SLEDrevs}. The real issue is to check (as is done below) whether the choices required for flat 4D geometries are stable against integrating out short-wavelength modes.}
\end{enumerate}

\subsubsection*{Reformulating the cosmological constant problem}

The existence of extra-dimensional solutions that allow flat 4D geometries to coexist with large 4D-lorentz-invariant energy densities does not in itself solve the cosmological constant problem. For different choices of brane properties there are also other solutions for which the on-brane geometries are curved.\footnote{de Sitter solutions to these equations are explicitly known \cite{6DdS}. This is interesting in its own right as a counter-example \cite{6Dnogonot} to no-go theorems for the existence of de Sitter solution in supergravity \cite{6Dnogo}.} One must re-ask the cosmological constant question in the 6D context: first identify which features of the branes are required for flat brane geometries, and then ask whether {\em these} choices are stable against integrating out high-energy degrees of freedom.

\medskip\noindent{\em A sufficient criterion for flat branes}

\medskip\noindent
At the classical level it turns out that there is a sufficient condition for the 4D brane geometries to be flat that is easy to state: they are flat provided the branes do not couple at all to the 6D dilaton \cite{6Dflatnophi} ({\em i.e.} that the functions $\cT_b$ and $\cA_b$ do not depend on $\phi$). To see why this is true, consider the Einstein and dilaton field equations arising from the action \pref{SSUSY},
\bea \label{EinsteinMN}
 \cR_{\ssM \ssN} + \partial_\ssM \phi\,\partial_\ssN \phi + \kappa_6^2 \, e^{-\phi} \, F_{\ssM\ssP} {F_\ssN}^\ssP  \qquad\qquad\qquad\qquad &&\nn\\
 -\frac12 \left( \frac{\kappa_6^2}4 \, e^{-\phi} F_{\ssP\ssQ} F^{\ssP \ssQ} - \frac{2 e_6^2}{\kappa_6^2} \, e^\phi\right) g_{\ssM \ssN} &=& - \kappa_6^2 \, \left( T_{\ssM \ssN} - \frac14 \, g_{\ssM\ssN} g^{\ssP \ssQ} T_{\ssP \ssQ} \right) \nn\\
 \Box \phi + \left( \frac{\kappa_6^2}4 \, e^{-\phi} F_{\ssP\ssQ} F^{\ssP \ssQ} - \frac{2 e_6^2}{\kappa_6^2} \, e^\phi\right) &=& -\kappa_6^2 \, \cJ \,,
\eea
where
\be
 T^{\ssM \ssN} = \sum_b \frac{2}{\sqrt{-g}} \; \frac{\delta S_b}{\delta g_{\ssM \ssN}} \; \delta^2(x)   \quad \hbox{and} \quad
 \cJ = \sum_b \frac{1}{\sqrt{-g}} \; \frac{\delta S_b}{\delta \phi} \; \delta^2(x) \,,
\ee
are the corresponding brane sources. For the cosmological constant our interest is in maximally symmetric 4D configurations, for which $\partial_\mu \phi = F_{\mu \ssN} = 0$ and so $\Box \phi = \nabla^2 \phi$, where $\nabla^2 = g^{mn} \nabla_m \nabla_n$ is the 2D laplacian in the extra dimensions.

These equations relate the near-brane derivatives of the bulk fields to the properties of the brane actions, as follows \cite{Cod2Matching}. Imagine integrating the dilaton field equation over an infinitesimal volume, $V_\epsilon$, that includes the position of the brane. Here $\epsilon$ is the radius of this volume, that is taken to zero after the integration is performed. On the right-hand side the delta function performs the integral, and the result vanishes if the brane does not couple to the dilaton, since $\cJ \propto \delta S_b/\delta \phi = 0$. For infinitesimal integration volume only the second-derivative term on the left-hand side gets a nonzero contribution, but is a total derivative and so can be written as a surface integral
\be
 \int_{V_\epsilon} \exd^2x \sqrt{-g} \; \nabla^2 \phi = \int_{V_\epsilon} \exd^2x \, \partial_\ssM \Bigl( \sqrt{-g} \; \nabla^\ssM \phi \Bigr) = \oint_{\partial V_\epsilon} \exd x \sqrt{-g} \; n_\ssM \nabla^\ssM \phi \,,
\ee
where $n_\ssM$ is the normal to the boundary, $\partial V_\epsilon$. The absence of brane couplings to $\phi$, $\delta S_b/\delta \phi = 0$, implies the radial derivative of $\phi$ must vanish in the near-brane limit. An identical argument applied to the 2D trace of the Einstein equations
\be \label{Einsteinmn}
 \int_{V_\epsilon} \exd^2x \sqrt{-g} \; g^{mn} \cR_{mn} = - \frac{\kappa_6^2}2 \int_{V_\epsilon} \exd^2x \sqrt{-g} \;  \left( g^{mn} T_{mn} - g^{\mu \nu} T_{\mu\nu} \right) \,,
\ee
relates the near-brane extra-dimensional geometry to the source stress-energy (which is not needed in what follows, but includes the relationship between conical defect angle and brane tension). As applied to the 4D trace of the Einstein equations it gives
\be \label{Einsteinmunu}
 \int_{V_\epsilon} \exd^2x \sqrt{-g} \; g^{\mu\nu} \cR_{\mu\nu} = \kappa_6^2 \,  \int_{V_\epsilon} \exd^2x \sqrt{-g} \; g^{mn} T_{mn} \,,
\ee
which, when using $\sqrt{-g} = \sqrt{- \hat g} \; e^{4W}$ and
\be \label{warpedR}
 g^{\mu\nu} \cR_{\mu\nu} = e^{-2W} \hat g^{\mu\nu} \hat R_{\mu\nu} + e^{-4W} \nabla^2 e^{4W} \,,
\ee
makes the near-brane limit of $n_\ssM \nabla^\ssM W$ proportional to $g^{mn} T_{mn}$ of the source.

One can now restrict to 4D matter that only lies in the brane directions, for which only the 4D components of $T_{\ssM\ssN}$ are nonzero, $T_{m\ssN} = 0$. For this kind of matter maximal symmetry also requires $T_{\mu\nu} = - \rho_{\rm vac} \, g_{\mu\nu}$, and so $T_{\ssM\ssN}$ drops out of the right-hand side of the 4D Einstein equations. Notice also that the vanishing of $g^{mn} T_{mn}$ implies the near-brane limit of $n_\ssM \nabla^\ssM W$ vanishes.

To learn what this implies for the on-brane curvature, $\hat R = \hat g^{\mu\nu} \hat R_{\mu\nu}$, we return to the 4D trace of the Einstein equations, but now integrate them over the entire extra dimensions {\em outside} of the small volumes, $X_2 = M_2 - \sum_b V_\epsilon$, that surround the sources, keeping track of the finite parts that previously vanished in the $\epsilon \to 0$ limit when integrating over $V_\epsilon$. Using the dilaton equation in the 4D trace of the Einstein equation we have,
\be
 \int_{X_2} \exd^2x \sqrt{-g} \; \left( g^{\mu\nu} \cR_{\mu\nu} + \frac12 \, \nabla^2 \phi \right) = 0 \,.
\ee
There is no contribution on the right-hand side because the delta functions all have support only inside $V_\epsilon$ and not outside. It follows, using eq.~\pref{warpedR} and the property $n_\ssM \nabla^\ssM \phi = 0$ (because $\delta S_b/\delta \phi = 0$), that
\bea
 \int_{X_2} \exd^2x \sqrt{- \hat g} \; e^{2W} \hat R &=& - \int_{X_2} \exd^2x \sqrt{- \hat g} \; \nabla^2 e^{4W} =  - \oint_{\partial X_2} \exd^2x \sqrt{- \hat g} \; n_\ssM \nabla^\ssM e^{4W} \nn\\
 &=& + \sum_b \oint_{\partial V_\epsilon} \exd^2x \sqrt{- \hat g} \; n_\ssM \nabla^\ssM e^{4W} = 0 \,,
\eea
where the final equality uses the boundary condition found above that $n_\ssM \nabla^\ssM W = 0$ on the boundaries of the small volumes, $V_\epsilon$.

The upshot is that there is a remarkably simple sufficient condition for the 4D on-brane geometry to be flat: the absence of brane couplings to the dilaton, $\delta S_b/\delta \phi = 0$, and the absence of off-brane stress energy, $g^{mn} T_{mn} = 0$. The next question is whether or not these conditions are stable under quantum corrections.

\medskip\noindent{\em Stability under quantum corrections}

\medskip\noindent
These two necessary criterion are remarkable in that they are precisely the kinds of criteria that can be preserved by loops of the heavy Standard Model particles (which are at the root of the problem), because these reside on the branes.

For example, having large tensions makes the branes like to be straight and so not bend into the off-brane directions. So any particles trapped on such branes will have stress energy only in the on-brane directions: $T_{m\ssN} = 0$. On-brane loops do not make the branes more likely to bend, because they typically do not reduce the brane tension.

Second, given there are initially no couplings between any on-brane particles and the bulk dilaton, such couplings are never generated by doing only loops of on-brane particles because these are functions only of the $\phi$-independent initial couplings.

These arguments are indeed borne out by explicit calculations \cite{BraneLoops}. But what of quantum corrections that also involve the extra-dimensional, bulk fields? These are potentially dangerous, because they can generate brane-$\phi$ couplings even if these are initially not present. They can do so because the brane does couple to the metric (and to the bulk gauge field) and these couple to the dilaton, so loops involving the metric and gauge field can generate nonzero couplings between branes and $\phi$.

How big are these? Here it happens that the supersymmetry of the bulk comes to the rescue. Explicit calculations \cite{BulkLoops, SUSYBLF} show that supersymmetry can allow the effective 4D vacuum energy to be of order the Kaluza-Klein scale, $m_\KK^4$. Bulk supersymmetry can help for two reasons. First, although loops involving various bulk fields can and do generate all possible kinds of brane-bulk interactions, most of these cancel once summed over the field content of a 6D supermultiplet. This occurs because the most dangerous loops are those involving very short wavelength modes, but such modes typically do not `know' that supersymmetry has broken because they are too short to see the boundary conditions at distant branes, where supersymmetry breaks.

Second, those quantum bulk contributions that do not cancel tend naturally to be very small. This is ultimately because an $N$-loop contribution is always proportional to a factor of $e^{2N\phi}$, which is a very small number. Each loop comes with a factor of $e^{2\phi}$ because this is the loop-counting parameter that controls the validity of the semiclassical approximation. The simplest way to see this is to rewrite the bulk action, \pref{SSUSY}, by redefining $g_{\ssM\ssN} = e^{-\phi} \hat g_{\ssM\ssN}$, in which case it takes the form
\be
 S_\ssB = - \int \exd^6 x \sqrt{-\hat g} \; e^{-2\phi} \,\cL(\hat g_{\ssM\ssN}, F_{\ssM\ssN}, \partial \phi) \,.
\ee
The point is that $\phi$ only appears undifferentiated in front of the entire action, making $e^{2\phi}$ appear in the same way as does $\hbar$ in the combination $S/\hbar$.

What makes $e^{2\phi}$ small is eq.~\pref{Lvsphi}, which states that $e^{2\phi} \propto 1/(ML)^4$, where $L$ is the physical size of the extra dimensions and $M$ is a TeV-sized mass scale (coming from $e_6/\kappa_6$). This means that a one-loop correction to a generic vacuum energy is of order
\be \label{rho1loop}
 \delta \rho_{\rm vac}(1 \; \hbox{loop}) \simeq \frac{k}{(4\pi)^2} \, M^4 \, e^{2\phi} \simeq \frac{k}{(4 \pi L^2)^2} \,,
\ee
where $k$ is a dimensionless, calculable number (more about which below). But this can be of order the observed cosmological constant when the extra dimensions are large: $1/L \simeq 10^{-2}$ eV. Higher bulk loops are suppressed by four more powers of $1/L$, and so are totally irrelevant. Unusually, the cosmological constant problem is only a one-loop issue for the bulk, because of the remarkably small bulk coupling.

Indeed, in certain circumstances the result can even be smaller than this because the branes sometimes fail to break supersymmetry at one loop \cite{SUSYBLF}. In this case the leading quantum contribution to the observed 4D cosmological constant turns out to involve one brane and one bulk loop, and so would be of order
\be \label{rho2loop}
 \delta \rho_{\rm vac}(1 \; \hbox{bulk} + 1 \; \hbox{brane loop}) \simeq \frac{k'}{(4\pi)^4} \, M^4 \, e^{2\phi} \simeq \frac{k'}{(4 \pi L)^4} \,,
\ee
where $k \simeq k'/(4 \pi)^2$ for a new dimensionless constant, $k'$.

Is this small enough? This depends on how large $L$ is. There are two phenomenological upper limits for $L$ that limit how low eqs.~\pref{rho1loop} and \pref{rho2loop} can be.
\begin{itemize}
\item {\em Tests of Newton's inverse-square law} provide the strongest constraints on the existence of extra dimensions in the sub-eV range. In the conventions used here, these presently give an upper limit \cite{Adelberger} $L < L_e \simeq 45 \; \mu$m, and so $1/L > 1/L_e \simeq 4 \times 10^{-3}$ eV, so $L < L_e$ is not inconsistent with \pref{rho1loop} or \pref{rho2loop} giving the observed dark energy density, \pref{rhovobs}, because of the loop factors of $4\pi$.
\item {\em Astrophysical energy-loss constraints} turn out to provide a more stringent limit on extra dimensions \cite{Raffelt} and their supersymmetric extensions \cite{SUSYraffelt, MSLED}. The most stringent bounds\footnote{There are other bounds \cite{PDG, StrongerBounds} on large dimensions that are stronger than these, but they are more model-dependent inasmuch as they rely on extra-dimensional particles decaying into visible particles (like photons). They can be evaded by constructions that provide more efficient channels for these particles to decay invisibly.} come from the cooling rates for red giant stars and supernovae, since these can be appreciably modified if radiation into the extra dimensions is too efficient a cooling mechanism.

    These constrain the extra-dimensional gravity coupling, $\kappa_6 < \kappa_\SN \simeq (10 \; \hbox{TeV})^{-2}$, but a constraint on $L$ is possible given the relation between 4D and 6D gravitational couplings:\footnote{We use here that the defect angle must be small: $|\beta - 1| \simeq \cO(\kappa_6^2 \cT_b) \ll 1$.} $\kappa_6^2 = 4 \pi \beta L^2 \kappa^2 \simeq 4 \pi L^2 \kappa^2$:
    \be
     \frac{1}{L} = \frac{\sqrt{4\pi} \; \kappa}{\kappa_6} > \frac{1}{L_\SN} = \frac{\sqrt{4\pi} \; \kappa}{\kappa_\SN} \simeq 0.15 \; \hbox{eV} \quad \hbox{or} \quad
     L < L_\SN \simeq 1.3 \; \mu \hbox{m} \,.
    \ee
    Using $L_\SN$ in eqs.~\pref{rho1loop} and \pref{rho2loop} gives the intriguing limits $\rho_{\rm vac}(1 \; \hbox{loop}) \simeq (0.04 \; \hbox{eV})^4$ and $\rho_{\rm vac}(2 \; \hbox{loop}) \simeq (0.01 \; \hbox{eV})^4$, which, given the cavalier treatment of order-unity factors, is very close to the observed density.\footnote{It turns out that more detailed calculations \cite{LoopPis} show that the constant $k'$ is further suppressed by powers of the defect angle at each brane and the brane-localized fluxes. If these are all of the same order, $\kappa^2 \cT_b \simeq 1/4\pi$, then the total additional suppression is of order $1/(4\pi)^3$, leading to the remarkably good estimate $\delta \rho_{\rm vac} \simeq 4\pi k''/(16 \pi^2 L_\SN)^4 \sim 4\pi k'' (9 \times 10^{-4} \, \hbox{eV})^4$.}
\end{itemize}
Two conclusions can be drawn from these estimates. First, given the absence of larger scales it is worth performing the bulk vacuum energy calculations more precisely to pin down the order-unity factors for comparison with the observed dark energy density. Second, it is worth pushing tests of Newton's inverse-square law down to the micron level since deviations (a crossover to an inverse 4th power \cite{Callin}) at these distances is a smoking gun for the entire extra-dimensional  framework.

In the end SLED relies partially on supersymmetry and partially on the classical scale invariance that extra-dimensional supergravities usually have automatically. But if scale invariance plays an important role, why isn't Weinberg's no-go theorem a problem? That is, why don't loops drive a runaway to the scale-invariant minimum? Weinberg's argument indeed applies, and in spades because the quantum corrections need not be scale invariant. But nothing in Weinberg's argument dictates the {\em size} of these corrections, and in SLED they are smaller than expected because of the bulk supersymmetry. It is this interplay between supersymmetry and scale invariance that provides the wiggle room in which to seek a real solution to the problem.

\subsubsection*{Opportunities and worries}

Many questions remain, but from the point of view of this school two questions loom (like Mt.~Blanc) above the rest: Is this a solution to the cosmological constant problem? Is this the way Nature works and, if so, how would we tell? Here is a personal snapshot of the observational opportunities and potential pitfalls of the extra-dimensional approach.

\medskip\noindent{\em Opportunities}

\medskip\noindent
Start with the opportunities. As advertised in the introduction, there are a variety of tests of this picture, as there must be for any proposal which deals with the `old' cosmological constant problem. If Nature really cares whether the cosmological constant is technically natural, and if supersymmetric large dimensions provide the solution, then expect to see:
\begin{itemize}
\item {\em Tests of Newton's inverse square law:} As described above the gravitational inverse-square law must cross over to an inverse 4th power law \cite{Callin} at distances tied to the observed dark-energy density (of order a micron). Although the crossover is to a power-law, unfortunately for distances, $r > L$, the leading deviations from the inverse square are exponentially suppressed by a factor $\exp(- m_\KK r)$. (The Kaluza-Klein scale, $m_\KK$, depends on the extra-dimensional shape, but for a near-spherical example $m^2_\KK = l(l+1)/L^2$ with $l = 0,1,\cdots$ and so the lowest nonzero mass satisfies $1/m_\KK = L/\sqrt2 \simeq 0.9 \; \mu$m if $L \simeq 1.3 \; \mu$m.)
\item {\em Signals in the LHC:} If the extra-dimensional Planck scale is of order 10 TeV then the Large Hadron Collider must be operating close to a regime where quantum gravity is important. Potential signals must include energy loss into the extra dimensions \cite{ADDpheno, MSLED, SLEDpheno}, although these would not yet have been expected to have been seen for $M_6 = \kappa_6^{-1/2} > 10$ TeV \cite{PDG, LHCconfs}. More detailed predictions depend on what the theory of quantum gravity at these scales really is. String theory provides the most precise framework in which to ask the question,\footnote{Unfortunately it is not yet known how to get the 6D chiral gauged supergravity from string theory, though attempts have been made \cite{From10D}.} and suggests \cite{LHCStrings} that all Standard Model particles are likely to be accompanied by a tower of string excitations, split by the string scale, $M_s$. The observable signals for this are likely to be similar to those for excited Kaluza Klein modes, for which searches are on-going \cite{PDG, LHCconfs}. If so, they should be much easier to see than is missing energy. Furthermore they should very soon be accessible if $M_6 \simeq 10$ TeV, since in string models $M_s < M_6$.
\item {\em Modifications to gravity at longer distances} may also be possible, if any new light degrees of freedom survive into the low-energy 4D theory with masses below the Kaluza-Klein scale. If sufficiently light these provide both opportunities and dangers when compared with the wealth of tests of General Relativity. The precise nature of the 4D spectrum and the associated 4D effective theory in these theories is not yet settled, but is under active study. Included in this unfinished story will be the prediction of whether dark energy is really just a cosmological constant ($w = -1$), or has some sort of time dependence. (At present writing, the best guess is that there are no new states lighter than the KK scale, so it will prove to be just a cosmological constant.)
\item {\em Neutrino physics} need not, but could, contain observable signals. The neutrino masses indicated by oscillation experiments are not that different from the Kaluza-Klein scale in these models, and much study \cite{XDnus} has gone into how on-brane states might mix with bulk fermions if these were present and were for some reason light. Although such mixing is not required by the cosmological constant, supersymmetry makes these models more plausible by providing a host of bulk fermions all of whom are required to be light because they are tied by supersymmetry to the massless graviton. Furthermore, the same back-reaction of the branes onto the bulk that helps with the cosmological constant also helps somewhat with the neutrino phenomenology \cite{SLEDnu}. If such mixing is present neutrino physics should contain potentially interesting evidence for a family of sterile neutrinos.
\end{itemize}

To these can also be added two somewhat weaker predictions that have already succeeded:
\begin{itemize}
\item {\em Supersymmetry and the absence of Standard Model super-partners:} If nothing else, supersymmetric large extra dimensions provide an unusual way for supersymmetry to be realized in Nature, broken at the electroweak scale. Because supersymmetry is nonlinearly realized in the Standard Model sector, a SUSY transformation on an ordinary particle (say, the electron) gives a two-particle state (the electron plus a gravitino) rather than the single-particle state (the selectron) that the linear realization of the MSSM assumes. Consequently there is no requirement to find evidence for such superpartners at the LHC.
\item {\em A vanilla Higgs:} Given that the quantum gravity scale is so low (about 10 TeV) and given that the Standard Model works so incredibly well, there is not a lot of room to engineer a fancy Higgs sector on the brane. So in all likelihood the Higgs sector on the brane is just described by the Standard Model itself, leading one to expect a vanilla Higgs (such as has recently been discovered at the LHC \cite{HiggsLHC}).
\end{itemize}

It is satisfying that these last predictions --- both of which were made before the LHC results --- have been largely vindicated. Indeed, if the rest of these predictions were equally well verified it would be hard to miss.

\medskip\noindent{\em Potential pitfalls}

\medskip\noindent
But is this really, at long last, a solution to the cosmological constant problem? As always, spectacular claims require spectacular evidence. In particular, how does the SLED proposal do when compared with the three criteria of \S\ref{ssec:Criteria}?

It does very well on criteria 1 and 2: it is among the extremely few theories for which quantum corrections can be incorporated, and it does so for an energy range that includes the weak-scale particles whose masses are at the root of the problem. Of course, this is not good enough in itself. After all, another theory which also satisfies criteria 1 and 2 is standard MSSM supersymmetry, if the supersymmetry breaking scale is chosen to lie in the sub-eV range. This theory is rightly rejected because of criterion 3: the light super-partners that are predicted are not seen.

So before declaring victory with SLED it is necessary to be absolutely sure that nothing else has been ruined in the low-energy theory. Most of particle physics should come out well, given that non-gravitational on-brane interactions can be simply those of the Standard Model. But an important ingredient that is as yet missing for this framework is a theory of dark matter, and it is important because this is one of our few indications that the Standard Model fails.

Another issue of concern asks: Why should the extra dimensions be so large in the first place? Understanding this requires understanding the physics that stabilizes the modulus $\phi_0$ of the extra dimensions. Some ways for doing so have been explored and as we have seen flux quantization is one of the options \cite{LargeDimsCurvature}. This is how extra dimensions are stabilized in SLED, although at present the extra dimensions are made large by inserting a very small number directly into the lagrangian (through the value of $\cA_b$). It would be preferable to only put a number of order 50 into the lagrangian and find that the extra dimensions acquire a radius that is the exponential of this. Interestingly, flux quantization can can generate exponentially large dimensions in this way, if the branes couple directly to the dilaton, $\phi$ \cite{LargeDimsCurvature}. But because of this coupling systems with exponentially large dimensions appear to be disjoint from those that give a very small cosmological constant on the branes. An understanding of how the 6D theory can arise from ultraviolet-complete higher-dimension models is likely to help understand the choices available.

In some ways the extra-dimensional framework does just what is required: it modifies only the gravitational sector of the theory, and does so at very low energies without doing damage to non-gravitational tests. It can do so because only gravity `sees' the extra dimensions because all Standard Model particles are confined to a brane. Furthermore, it does so specifically for slow rather than fast process, as was argued earlier to be necessary. The slow/fast response arises because it is the way the extra dimensions back-react to the presence of the brane that underlies the small 4D curvature; after all, in the simplest models it is the conical curvature of the extra dimensions that cancels the brane tensions in the effective 4D cosmological constant \cite{NonSUSYRugby}. But the extra dimensions do not have time to respond to changes on the brane that happen faster than the extra-dimensional light-crossing time: $\tau \simeq L/c \simeq 3 \times 10^{-15}$ sec.

So the greatest worry at present is that the same cancellations that make the vacuum energy gravitate less than expected also do the same for static mass distributions. That is, does extra-dimensional back-reaction unacceptably modify how macroscopic and slowly moving objects gravitate? This involves exploring solutions on the brane that are not maximally symmetric; an analysis of which is in progress.

A linearized treatment gives a preliminary picture of what may be going on, however  \cite{Linear6DGrav, Linear6DSugra}. Imagine that two small stress energy distributions, $(T_{i})^{\ssM\ssN}$ with $i = 1,2$, interact through the exchange of a graviton in $D$ dimensions. Single-graviton exchange produces an interaction amplitude that is proportional to
\be \label{linearizedgrav}
 V_{12} \propto 2 \kappa^2_\ssD  \left[ (T_{1})^{\ssM\ssN} (T_{2})_{\ssM\ssN} - \frac{1}{D-2} \, (T_{1})^\ssM_\ssM (T_{2})^\ssN_\ssN \right] \,,
\ee
where, as usual, $\kappa^2_\ssD$ is the gravitational coupling constant that appears in the $D$-dimensional Einstein-Hilbert action. In the usual $D=4$ case this gives
\be
 V_{12} \propto \kappa^2_4 \, m_{1} m_{2} \,,
\ee
for the interactions of two non-relativistic objects (for which $T_{i}^{\mu\nu} = m_{i} \, \delta^\mu_0 \, \delta^\nu_0$), and once the constants of proportionality are included this gives the standard result $\kappa^2_4 = 8 \pi G_\ssN$ relating $\kappa_4$ to Newton's constant. For scattering of a massive object --- $(T_{1})^{\mu\nu} = m_{1} \delta^\mu_0 \delta^\nu_0$ --- with a photon --- for which $(T_{2})^\mu_\mu = 0$ ---  the interaction amplitude is instead
\be
 V_{12} \propto 2 \, \kappa^2_4  m_{1} T^{00}_{2} = 2 G_\ssN m_{1} T^{00}_{(2)} \,,
\ee
which displays a famous factor of 2 in the prediction for the bending of light about a massive source in General Relativity.

As applied to the scattering of particles having only 4D stress energy --- $(T_{i})^{m\ssN} = 0$ --- in 6D, eq.~\pref{linearizedgrav} implies
\be \label{linearizedgrav6}
 V_{12} \propto 2 \kappa^2_6  \left[ (T_{1})^{\mu\nu} (T_{2})_{\mu\nu} - \frac{1}{4} \, (T_{1})^\mu_\mu (T_{2})^\nu_\nu \right] \,.
\ee
Notice that this captures several features of the above discussions, and in particular vanishes if $T_{i}^{\mu\nu} = - \cT_{i} \, g^{\mu\nu}$ is the Lorentz-invariant tension (or vacuum energy) localized on a brane.

For interactions at distances very large compared with the size of the extra dimensions $V_{12}$ receives contributions only from those states that are massless in 4D. The result for 6D graviton exchange differs from the result for 4D graviton exchange in this regime because of the contribution of massless 4D scalars (one of which starts life in 6D as an overall rescaling of the extra-dimensional metric, and so couples to the trace of the 4D stress energy) \cite{Linear6DGrav}:
\bea \label{4Dvs6Dforce}
 V_{12} &\propto& 2 \kappa^2_6  \left[ (T_{1})^{\mu\nu} (T_{2})_{\mu\nu} - \frac{1}{4} \, (T_{1})^\mu_\mu (T_{2})^\nu_\nu \right] \nn\\
 &=& 2 \kappa^2_6  \left[ (T_{1})^{\mu\nu} (T_{2})_{\mu\nu} - \frac{1}{2} \, (T_{1})^\mu_\mu (T_{2})^\nu_\nu \right] + 2 \kappa^2_6  \left[ \sum_i c_i \, (T_{1})^\mu_\mu (T_{2})^\nu_\nu \right] \,,
\eea
where $\sum_i c_i = \frac14$. If these 4D scalars get a nonzero mass, of order $m_0$, the second term on the second line of \pref{4Dvs6Dforce} drops out at distances longer than $m_0^{-1}$. So the interaction looks 6-dimensional if sampled over ranges smaller than than $m_0^{-1}$, but it agrees with the standard 4D result for ranges larger than $m_0^{-1}$.

If these scalars get a mass at or just below the KK scale, $m_0 \simeq \delta_b m_\KK$, (where $\delta_b \simeq \kappa^2 \cT_b$ is of order the brane defect angle, as seems to be the case) then at the linearized level we therefore expect gravitational interactions to look 6-dimensional (and so predict vanishing interactions for a 4D vacuum-energy source) out to ranges of order the size of the extra dimensions. But gravitational interactions should start to look 4-dimensional at distances larger than this (and so the vacuum energy should start to stop cancelling and have influences over these scales). This all looks very good because it would reproduce 4 dimensional gravity over the scales for which it has been tested ({\em i.e.} distance longer than the KK -- or micron -- scale) while preserving the cancellation of the vacuum energy over distances shorter than these same scales -- which are the ones that are at the root of the cosmological constant problem.

What is not yet clear is how this goes through beyond the linear regime. After all, the exact classical solutions discussed above for maximally symmetric branes explore the nonlinear regime, and show that the vacuum energy continues to cancel there. So one wonders whether the Newtonian interactions of massive point masses might also deviate from the 4D picture at this nonlinear level. Calculations to resolve this issue and establish the long-distance form of gravitational interactions at the nonlinear level are in progress.

\section{Summary}
\label{sec:summary}

So after this long journey, what is the bottom line? These lectures have tried to make the following points.

\begin{enumerate}
\item It has been argued that technical naturalness matters, and is a natural consequence of the modern meta-understanding of what we really do when we write down a classical lagrangian to describe a particular system. What we construct is just an effective field theory, designed to capture the important physics relevant to some range of energy scales. Because there can be a different effective theory for every new range of scales, within this context it should be possible to ask why a parameter is small at any scale we choose. A technically natural understanding of a small parameter asks why it is small in the ultraviolet-complete theory at very high energies, and then separately asks why it stays small as one integrates out the lower-energy modes.

    Although we may not understand the answer of the first question until we get access to very high energies, we should be able to understand the second part even at low energies (and this is what makes the criterion of technical naturalness useful). Although it is logically consistent to use theories that are not technically natural, to my knowledge all of the myriad well-understood hierarchies of scale between cosmology and the frontiers of particle physics have a technically natural understanding, so it is a radical proposal to assume that the cosmological constant should be different.
\item Technical naturalness is relevant to the `old' cosmological constant problem, which asks why the energy of the vacuum (which we believe should be very large) doesn't appear to gravitate. This is independent of the various `new' cosmological constant problems, that take the `old' problem as being somehow solved to give zero vacuum energy, and then ask how one comes up with the observed accelerated expansion of the Universe.
\item It is argued that it is misleading to track ultraviolet divergences when asking if a theory is technically natural, because the one thing we know about regularization schemes is that they cancel out of observables once these are expressed in terms of renormalized quantities. What is relevant is how renormalized parameters depend on renormalized masses. This makes the difference between theory and experiment for the cosmological constant a factor of order $10^{50}$ and not of order $10^{122}$. The electron alone is a big problem, since its predicted contribution to the vacuum energy seems to be more than $10^{30}$ times too large.
\item Three criteria were given against which to compare candidate solutions to the `old' cosmological constant problem. The theory must ($a$) work at at the quantum level; ($b$) make sense up to an ultraviolet scale that is larger than the sub-eV scale of the observed dark-energy density; and ($c$) not ruin anything else that we know about physics at the relevant scales. It was argued that none of the popular models on the market satisfy all three.
\item It is argued that it is nonetheless too early to despair about solving the problem, and that two supersymmetric large extra dimensions provide a promising direction to explore. It is a promising direction because extra dimensions break the direct link between a 4D Lorentz-invariant vacuum energy and a large 4D curvature. Furthermore, supersymmetry in the extra dimensions combines scale invariance and supersymmetry in a way that partially exploits the advantages of both for suppressing the cosmological constant (and evading Weinberg's no-go result).

    The cosmological constant problem has been reposed in the higher dimensional context. The criteria for having small 4D curvatures has been identified, and loop corrections to this criteria have been investigated and can be smaller than anticipated. Particles can have masses $M \gg m_\KK$ on the branes, and yet the loop-corrected 4D curvature can be as small as what would be expected for a vacuum energy of order $m_\KK^4$. The picture that emerges is a very supersymmetric gravity sector coupled to a non-supersymmetric particle-physics sector. In particular no supersymmetric partners to ordinary particles are expected at the LHC.

    If this is the way the world works we will soon know: besides the absence of superpartners at the LHC we must see deviations from Newton's inverse-square law at micron sizes and likely also see new physics at the LHC (and possibly for neutrinos). Conversely, it is easily falsifiable, such as if no deviations from Newton`s inverse square law is found over sum-micron distances. One would also be uncomfortable if no evidence for quantum gravity (string excited states, missing energy) were to be found at the full-energy LHC.
\end{enumerate}

In the end the cosmological constant problem is a message of hope and not despair. By being large, the Universe is telling us that new physics is just around the corner, since this is required by {\em any} real solution to the problem. So far the search has been hard and unsuccessful, but with luck the interplay between cosmology, experiments at TeV scales and tests of gravity will soon teach us what is really going on. Let us hope!

\section*{Acknowledgements}
I would like to thank the organizers of this school for their kind invitation to present these lectures in such pleasant environs. I have special thanks for several superb student/colleagues --- Yashar Aghababaie, Ross Diener, Doug Hoover, Leo van Nierop and Matt Williams --- and talented collaborators --- Jim Cline, Susha Parameswaran, Fernando Quevedo, Alberto Salvio, Gianmassimo Tasinato and Ivonne Zavala --- for their indulgence over many years and their help in understanding the many thorny issues raised by the cosmological constant problem. My research has been supported in part by the Natural Sciences and Engineering Research Council of Canada. Research at the Perimeter Institute is supported in part by the Government of Canada through Industry Canada, and by the Province of Ontario through the Ministry of Research and Information.

\end{document}